\documentclass[pra,onecolumn,superscriptaddress]{revtex4-2}

\usepackage{amsmath}
\usepackage[all]{xy}
\usepackage{amssymb}
\usepackage{graphicx}
\usepackage{subfigure}
\usepackage{amsthm}
\usepackage{xcolor}
\usepackage[colorlinks, linkcolor=red, anchorcolor=blue, citecolor=green]{hyperref}

\newcommand\Acal{\mathcal A}

\newcommand\Ocal{\mathcal O}

\newcommand\Id{\operatorname{Id}}
\newcommand\rk{\operatorname{rank}}
\newcommand\BQP{\operatorname{BQP}}
\newcommand\QMA{\operatorname{QMA}}
\newcommand\spn{\operatorname{span}}
\DeclareMathOperator{\tr}{tr}

\newtheorem{theorem}{Theorem}

\newtheorem{example}{Example}

\begin{document}

\title[Quantum Topological Analysis on Digraphs]{Quantum Topological Analysis on Digraphs}

\author{Muchun Yang} \email[]{yangmuchun21@mails.ucas.edu.cn}
\affiliation{Institute of Physics, Beijing National Laboratory for Condensed
  Matter Physics,\\Chinese Academy of Sciences, Beijing 100190, China}
\affiliation{School of Physical Sciences, University of Chinese Academy of
  Sciences, Beijing 100049, China}
\affiliation{Department of Physics, Tsinghua University, Beijing 100084, China}

\author{D. L. Zhou} \email[]{zhoudl72@iphy.ac.cn}
\affiliation{Institute of Physics, Beijing National Laboratory for Condensed
  Matter Physics,\\Chinese Academy of Sciences, Beijing 100190, China}
\affiliation{School of Physical Sciences, University of Chinese Academy of
  Sciences, Beijing 100049, China}

\author{Yunpeng Zi}\email[]{{ziypmath\_2024@email.sdu.edu.cn}}
\affiliation{School of Mathematics, Shandong University, Jinan, Shandong 250100, China}

\date{\today}

\begin{abstract}
Quantum algorithms for topological data analysis provide significant advantages over the best known classical algorithms. Unlike previous work on simplicial complexes built from point clouds, path homology on digraphs is defined for directed graphs and provides a natural topological framework for analyzing data with intrinsic directional structures. Path homology has become an emerging area in Topological Data Analysis (TDA), attracting increasing attention in recent years. We propose a quantum algorithm for path homology on digraphs that offers a significant advantage over the best known classical algorithms. We design a universal encoding protocol for the paths and boundary operators of digraphs on quantum systems. We prove a property of path homology that provides the theoretical guarantee for the algorithm. The speedup of the quantum algorithm for path homology depends on input-access assumptions. The exponential speedup arises when the path space can be efficiently accessed, while for standard digraph input the algorithm provides polynomial speedup.
\end{abstract}           
\maketitle
\section{Introduction}

Topological Data Analysis (TDA) is a mathematical field that focuses on the `shapes' of the data. By applying the topological tools, such as simplicial homology and knot theory, it is able to analyze complex high-dimensional data and identify key features of the data which may be difficult to extract using traditional methods.
A quantum algorithm for TDA is first proposed by Lloyd, Garnerone, and Zanardi (LGZ)~\cite{Lloyd2016}, and it provides significant advantage over the best classical algorithm. 
The foundations of the LGZ algorithm are linear-algebraic quantum machine learning (QML) algorithms~\cite{Biamonte2017, PhysRevLett.103.150502, PhysRevLett.122.040504, Lloyd2014, PhysRevLett.113.130503, Aaronson2015, Beer2020, Havlíček2019}, which are one of the most rapidly growing fields in the research of quantum computing. 
Recent studies on quantum algorithms for TDA have achieved notable theoretical progress~\cite{Cade2024complexityof, Hayakawa2022quantumalgorithm, gunn2019reviewquantumalgorithmbetti, ubaru2021quantumtopologicaldataanalysis, mcardle2022streamlinedquantumalgorithmtopological, PRXQuantum.5.010319, PRXQuantum.4.040349, PhysRevA.110.042616,George_2019,PhysRevA.106.022407,Ameneyro2024,Gyurik2022towardsquantum,Nghiem2025quantumalgorithm,Hayakawa2026quantumwalks,gvys-hl8h,schmidhuber2025quantumalgorithmkhovanovhomology,nghiem2026hybridquantumclassicalframeworkbetti,liu2025bettinumberspersistencediagrams,PhysRevApplied.23.054054,yamauchi2025quantumspectroscopytopologicaldynamics,parzygnat2026quantumencodingspreservepersistent,zucchini2024newquantumcomputationalsetup,m6nc-ypl7,nghiem2025quantumtopologicaldataanalysis,Vlasic_2023,Nghiem2025quantumalgorithm}. The LGZ algorithm has also been implemented experimentally on optical platform~\cite{Huang:18} and noisy quantum simulator~\cite{ICLR2024_9675b27e}.

The complexity of the LGZ algorithm is analyzed in Ref.~\cite{PRXQuantum.4.040349}. It shows that exact Betti-number computation is $\#\mathrm{P}$-hard, while multiplicative-error approximation is $\mathrm{NP}$-hard. Under ordinary vertex-and-edge input, the quantum advantage is therefore at most polynomial, whereas direct specification of simplices can enable an exponential advantage. A complementary resource analysis in Ref.~\cite{PRXQuantum.5.010319} further shows that the practical advantage depends strongly on the error criterion, the scaling of the Betti number, and the costs of state preparation and spectral projection.
Recent complexity-theoretic results have clarified important limitations of quantum Betti number computation in the simplicial setting. In particular, Crichigno and Kohler strengthened earlier NP-hardness results by showing that clique homology is $\#\BQP$-hard in the exact counting setting and $\QMA_1$-hard in the decision setting with exponentially fine precision, suggesting that efficient quantum algorithms are unlikely for generic worst-case instances~\cite{Crichigno2024}. By contrast, Lee and Nghiem showed that stronger input assumptions can still enable quantum advantages~\cite{lee2025newaspectsquantumtopological}.

Path homology on digraphs, first introduced and named by Alexander Grigor'yan, Yong Lin, Yuri Muranov and Shing-Tung Yau in 2012~\cite{grigor2012homologies}, is an emerging area in TDA that has attracted increasing attention in recent years.

It is a homology theory for describing topological properties of directed graphs, where the direction of edges carries essential structural information that cannot be fully captured by ordinary simplicial homology. Unlike simplicial homology, which is built from unordered simplices, path homology is based on ordered paths in a digraph. Therefore, it provides a natural framework for studying topological features of directed data.

In recent years, researchers have developed the homotopy theory based on the GLMY homology~\cite{grigor2014homotopy}, the GLMY cohomology theory~\cite{grigor2015cohomology}, the persistent GLMY homology~\cite{chowdhury2018persistent}, the analytic and Reidemeister torsions of digraphs~\cite{grigor2024analytic} and the iterated integral theory~\cite{yau2026algebra}. Moreover the GLMY homology theory has been generalized to describe the topological properties of multigraphs and Quivers in~\cite{GrigoryanMuranovVershininYau+2018+1319+1337}, weighted digraphs in~\cite{lin2019weighted} and weighted hypergraphs in~\cite{muranov2023homology}. The GLMY homology has also shown potential in the applied mathematics, for example the applications in deep feedback networks in~\cite{chowdhury2019path}, temporal networks in~\cite{chowdhury2022path}, molecular and materials sciences in~\cite{chen2023path}. However in the studies and applications of this theory, the efficient computation of GLMY homology has long remained a major challenge for mathematicians. Some attempts have been made to accelerate the computation~\cite{dey2022efficient,burfitt2024inductive}. However, the best known algorithm for computing $k$-th GLMY homology was the one given in~\cite{chowdhury2018persistent} which has time complexity $\Ocal(n^{3+3k})$ providing $n$ the number of vertices in the digraph.

Unlike simplicial homology, GLMY homology has two distinctive features. First, the vertices in a path are ordered, which affects the form of the boundary operator. Second, there is no canonical choice of generators for the chain group; in particular, these generators need not be orthogonal. Because of these features, the LGZ algorithm cannot be applied directly to GLMY homology.

Here we propose a quantum algorithm for the GLMY homology on digraphs. The framework incorporates three components: preparation of the chain-space input state, coherent implementation of the projected Dirac or embedded Hodge-Laplacian operator, and phase estimation of its spectrum. We design a quantum encoding of ordered paths and an implementation algorithm of the path boundary map. 
We prove that the embedded Hodge Laplacian can be obtained by projecting the total path Dirac operator onto the embedded chain space between successive Dirac actions. This projected-Dirac construction avoids explicitly constructing or inverting the Gram matrix associated with a non-orthogonal coordinate representation.
The complexity of the proposed algorithm depends crucially on the input assumptions, which determine the efficiency of preparing states supported on the path space. Accordingly, the exponential speedup arises in regimes with more direct access to the path data, while in standard digraph-input assumptions the algorithm is limited by state preparations and provides polynomial speedup.

\section{GLMY Homology: Definitions and Computations}
A digraph $G=(V,E)$ is a tuple where $V = \{1,\cdots,n \}$ is the set of vertices, and $E$ is a finite set of ordered pairs $(v_i\to v_j)$ with $v_i,v_j\in V$ and $v_i\neq v_j$, called the set of arrows or the set of directed edges. For simplicity, we assume that there are no loops in the digraph in the following discussion, i.e. there is no sequence of arrows $v_0\to v_1,v_1\to v_2,\cdots, v_{k-1}\to v_k$ where $k$ is a positive integer such that $v_0=v_k$.

\begin{figure*}
\subfigure[]{
\raisebox{5mm}{
        \includegraphics[width=4cm]{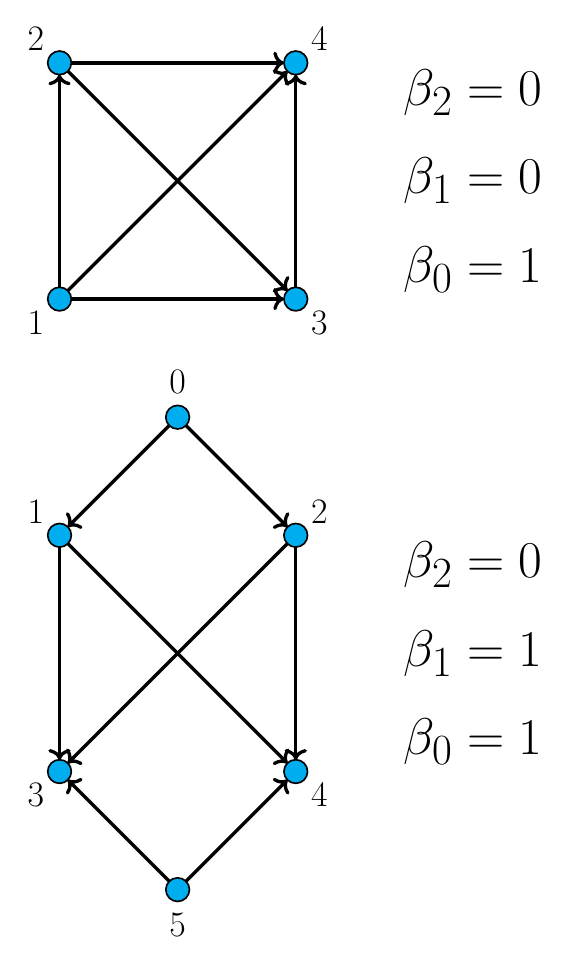}}
        }
\hspace{0mm}
\subfigure[]{
\includegraphics
[width=13.0cm]
{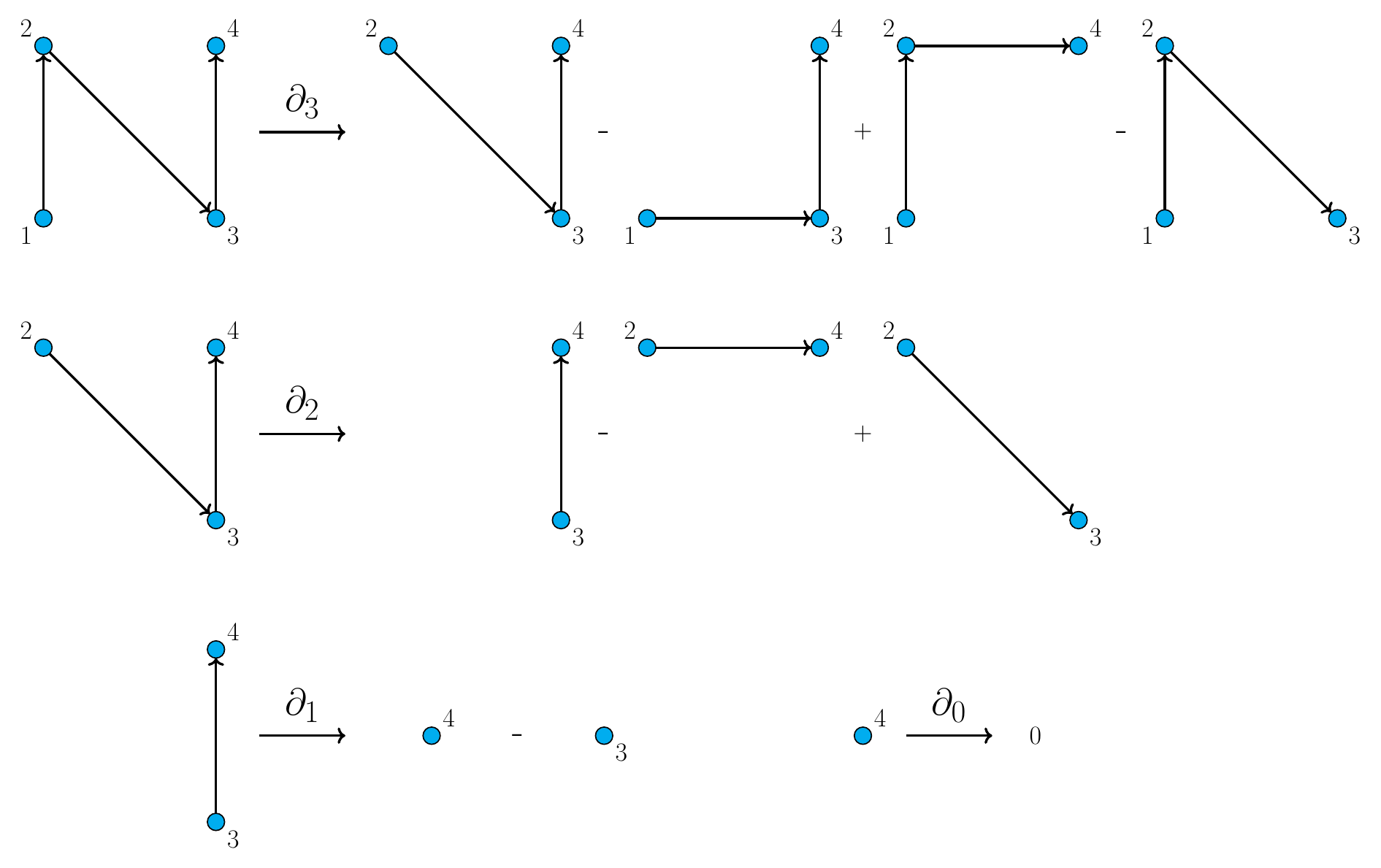}
}
\caption{
The illustration of digraph of GLMY homology and the boundary operator $\partial$ for GLMY homology. (a) Two digraphs with $n=4$ and $n=6$ and their Betti numbers. (b) The illustration of the boundary operator $\partial$ acting on paths.}
\label{fig1}
\end{figure*}

\subsection{Definitions}
An \textit{elementary $k$-path} is an ordered chain of vertices denoted as $v_0\cdots v_k$, where the number $k$ is called the length of the path. An elementary $k$-path $v_0\cdots v_k$ is called \textit{regular} if and only if $v_i\neq v_{i+1}$ for all $i\in\{0,\cdots,k-1\}$, and irregular otherwise. The space spanned by all regular elementary $k$-paths of length $k$ is denoted by $\Lambda_k$. If a regular elementary $k$-path satisfies $(v_i\to v_{i+1})\in E$ for every adjacent pair, then it is called \textit{allowed}. The space spanned by all allowed elementary $k$-paths in $G$ is denoted by $\Acal_k$.
The boundary map $\partial_k:\Lambda_k\to \Lambda_{k-1}$, which is given on each elementary $k$-path and extended linearly, is defined by
\begin{equation}\label{eq:boundary-path}
    \partial_k(v_0\cdots v_k)=\sum_{i=0}^k(-1)^iv_0\cdots \hat{v}_i\cdots v_k,
\end{equation}
where $\hat{v}_i$ means the vertex $v_i$ is removed from the path. We call $\partial_k$ the $k$-th boundary map. The illustrations of the paths on a digraph and the boundary map are shown in Fig.~\ref{fig1}.

A standard approach for constructing the GLMY Homology is by taking the chain complex as the vector space $\Omega_k = \{x\in \Acal_k: \partial_k x \in \Acal_{k-1}\}$ and restricting the boundary maps $\partial_k$ to the space $\Omega_k$, denoted as $\partial^{\Omega}_k$. It is clear that the spaces $(\Omega_k,\partial^{\Omega}_k)$ constitute a long chain with $\partial^{\Omega}_{k-1}\circ\partial^{\Omega}_k=0$,
\begin{align}\label{simplicial_chain_Omega}
    \Omega_k \stackrel{\partial^{\Omega}_k}{\longrightarrow} \Omega_{k-1} \stackrel{\partial^{\Omega}_{k-1}}{\longrightarrow} \dots \stackrel{\partial^{\Omega}_{1}}{\longrightarrow} \Omega_0 \stackrel{\partial^{\Omega}_{0}}{\longrightarrow} 0.
\end{align}
And the \textit{$k$-th GLMY homology group} $H_k(G)$ is defined as $\text{Ker} \partial^{\Omega}_k/\text{Im} \partial^{\Omega}_{k+1}$.
Practically, the computation of $\Omega_k$ and $\partial^{\Omega}_k$ could be difficult in classical computations
since it is based on solving linear equations which suffers from high computational complexity.

There is an alternative definition of GLMY homology which is known as the \textit{embedded homology group} and proved to be isomorphic to the GLMY homology~\cite{bressan2019embedded}.
Let 
\begin{align}
    \Gamma_k := \Acal_k + \partial_{k+1}\Acal_{k+1}.
\end{align}

The inclusion
$\iota_k:\Gamma_k\subset \Lambda_k$ is an embedding, therefore we take the boundary map $\partial^{\Gamma}_k$ as the restriction of $\partial_k$ on the space $\Gamma_k$. Clearly this boundary map satisfies the equation
\begin{equation}\label{Eq:partGam-Part}
    \iota_{k-1}\circ\partial_k^{\Gamma}=\partial_k\circ \iota_k.
\end{equation}
There is a chain
\begin{align}\label{simplicial_chain_Gamma}
    \Gamma_k \stackrel{\partial^{\Gamma}_k}{\longrightarrow} \Gamma_{k-1} \stackrel{\partial^{\Gamma}_{k-1}}{\longrightarrow} \dots \stackrel{\partial^{\Gamma}_{1}}{\longrightarrow} \Gamma_0 \stackrel{\partial^{\Gamma}_{0}}{\longrightarrow} 0.
\end{align}
The $k$-th embedded homology $\mathbf{H}_k$ is defined as the quotient group, 
\begin{align}
    \text{Ker } \partial^{\Gamma}_k / \text{Im } \partial^{\Gamma}_{k+1}.
\end{align}
The dimensions of the homology groups $\mathbf{H}_k$ are important topological invariants, which is named as the \textit{$k$-th Betti number $\beta_k$ of the GLMY homology of $G$}. Betti numbers are also the invariants widely used in the topological data analysis.

\subsection{The Hodge Laplacian Operator} \label{sec:hodge-lapl-oper}

There exists a naturally defined inner product  $\langle -,-\rangle$ on the space $\Lambda_k$, for any two elementary $k$-paths $v_0\cdots v_k$ and $u_0\cdots u_k$, $\langle v_0\cdots v_k,u_0\cdots u_k\rangle$ equals to 1 if and only if $v_0=u_0, v_1=u_1,\cdots,v_k=u_k$ and 0 otherwise.
The subspace $\Gamma_k$ is equipped with the inner product induced from $\Lambda_k$, denoted by $\langle-,-\rangle_{\Gamma}$, making the inclusion $\iota_k:\Gamma_k\hookrightarrow\Lambda_k$ an isometry onto its image.
The dual maps of $\partial_k$ and $\partial^{\Gamma}_k$, with respect to the inner products $\langle -,-\rangle$ on $\Lambda_k$ and $\langle -,-\rangle_{\Gamma}$ on $\Gamma_k$, are denoted by $\partial_k^{*}:\Lambda_{k-1}\to \Lambda_k$ and $(\partial^{\Gamma}_k)^{*}:\Gamma_{k-1}\to \Gamma_k$, respectively, and are defined by

\begin{equation}\label{dual_definition}
    \begin{aligned}
        &\langle\partial_{k} x,y\rangle&=&\ \langle x,(\partial_k)^{*}y\rangle,\\
    &\langle\partial^{\Gamma}_{k} x',y'\rangle_{\Gamma}&=&\ \langle x',(\partial^{\Gamma}_k)^{*}y'\rangle_{\Gamma},
    \end{aligned}
\end{equation}
where 
$x\in \Lambda_{k},\ x'\in\Gamma_k, y\in\Lambda_{k-1}$ and $y'\in \Gamma_{k-1}$
are arbitrary $k$-paths and $(k-1)$-paths. The relation between these operators is stated in the following theorem, whose proof is given in Appendix~\ref{proof:dual-commute}.
\begin{theorem}\label{thm:dual-commute}
    Let $\pi_k:\Lambda_k\to\Gamma_k$ be the orthogonal projection onto $\Gamma_k$ with respect to the inner product $\langle-, -\rangle$.
    The dual maps $\partial_k^{*}$ and  $(\partial^{\Gamma}_k)^{*}$ satisfy the relation that
    \begin{equation}\label{eq:dual-commute}
        (\partial_k^{\Gamma})^{*}=\pi_k\circ\partial_k^{*}\circ \iota_{k-1},
    \end{equation}
    i.e. the following diagram commutes,
    \begin{equation}
        \xymatrix{
            \Gamma_k\ar[d]^{\iota_k}&\Gamma_{k-1}\ar[l]^{(\partial_k^{\Gamma})^{*}}\ar[d]^{\iota_{k-1}}\\
            \Lambda_k \ar@/^/[u]^{\pi_k}&\Lambda_{k-1}\ar[l]^{\partial_k^{*}}.}
    \end{equation}
    Equivalently, the image of the restricted map $\partial_k^{*}\circ \iota_{k-1}$ projected onto $\Gamma_k$ coincides with the embedding of the $\Gamma$-dual.
\end{theorem}

The $k$-th \textit{total Hodge-Laplacian operator} 
$\Delta_k:\Lambda_k\to \Lambda_k$
with respect to the inner product $\langle -,-\rangle$ and the $k$-th \textit{Hodge-Laplacian operator} 
$\Delta^{\Gamma}_k:\Gamma_k\to \Gamma_k$
with respect to the inner product $\langle -,-\rangle_{\Gamma}$ is given as,
\begin{align}\label{Eq_Delta_k}
    &\Delta_k=(\partial_k)^{*}\partial_k+\partial_{k+1}(\partial_{k+1})^{*},
\end{align}
\begin{align}\label{dual_matrix}
&\Delta^{\Gamma}_k=(\partial^{\Gamma}_k)^{*}\partial^{\Gamma}_k+\partial^{\Gamma}_{k+1}(\partial^{\Gamma}_{k+1})^{*}.
\end{align}
It is well known~\cite{grigor2024analytic} that the $k$-th Betti number of the GLMY homology group of $G$ is computed as 
\begin{align}
  \label{eq:2}
    \beta_k =\text{dim}\ \text{Ker} \partial^{\Gamma}_k / \text{Im} \partial^{\Gamma}_{k+1} =\text{dim}\ \text{Ker}  \Delta^{\Gamma}_k.
\end{align}
In Appendix \ref{examples_classical_computation_GLMY}, we provide two examples of the classical computation of GLMY Homology.
More generally, the space $\Gamma_k$ could be factorized as the direct sum below called the \textit{Hodge decomposition}.
\begin{equation}
    \Gamma_k=\text{Ker} \Delta^{\Gamma}_k\oplus \partial^{\Gamma}_{k+1}\Gamma_{k+1}\oplus (\partial^{\Gamma}_{k})^{\ast}\Gamma_{k-1}.
\end{equation}

The following structural statement is the key operator reduction used in the quantum algorithm. Let
\begin{equation}
    \Lambda=\bigoplus_{k=0}^{d}\Lambda_k,
    \qquad
    \Gamma=\bigoplus_{k=0}^{d}\Gamma_k,
    \qquad
    \Pi=\bigoplus_{k=0}^{d}\pi_k,
\end{equation}
where $\pi_k:\Lambda_k\to\Gamma_k$ is the orthogonal projection with respect to the inner product introduced above. Define the ambient total boundary and ambient path Dirac operator by
\begin{equation}
    \partial=\bigoplus_{k=0}^{d}\partial_k,
    \qquad
    \partial^\dagger=\bigoplus_{k=0}^{d}\partial_k^{\dagger},
    \qquad
    B=\partial+\partial^\dagger
\end{equation}
Here $\partial_k^{\dagger}=(\partial_k)^{*}$ is the dual operator of $\partial_k$ on $\Lambda_k$. The operator $B$ is defined on the ambient space $\Lambda$, and its action is described purely by deleting or inserting positions in an ordered path. The embedded GLMY chain space is imposed by the projection $\Pi$. Thus the relevant Dirac operator on the embedded chain space is not the unrestricted $B$, but the projected operator
\begin{equation}
    B_\Gamma := \Pi B\Pi ,
\end{equation}
regarded as an operator on $\Gamma$. The theorem below states that this projected Dirac operator is exactly the Dirac operator of the embedded GLMY chain complex, and hence its square is the embedded Hodge Laplacian.

\begin{theorem}
\label{thm:total-laplacian}
On the embedded chain space $\Gamma$, the projected Dirac operator satisfies
\begin{equation}
    B_\Gamma=\Pi B\Pi
    =\partial^\Gamma+(\partial^\Gamma)^{*},
\end{equation}
where $\partial^\Gamma=\bigoplus_k\partial_k^\Gamma$ and the adjoint is taken with respect to the inner product inherited by $\Gamma$. Consequently,
\begin{equation}
    \Delta^\Gamma
    =B_\Gamma^2
    =\Pi B\Pi B\Pi .
\label{eq:projected-dirac-laplacian}
\end{equation}
In particular,
\begin{equation}
    \dim\ker\Delta_k^\Gamma=\beta_k .
\end{equation}
Thus the zero-eigenvalue sector relevant for the Betti number can be accessed by projecting the ambient path Dirac operator onto $\Gamma$, without explicitly constructing the Gram-matrix representation of $\Delta_k^\Gamma$ from a non-orthogonal generating set.
\end{theorem}
The proof is given in Appendix~\ref{proof:total-laplacian}.

To make the distinction from a direct coordinate construction explicit, let the columns of $E_k$ form a linearly independent basis of $\Gamma_k$ in the ambient orthonormal path basis, and set $N_k=E_k^\dagger E_k$. If $D_k$ is the ambient boundary matrix, then the coordinate matrix of the restricted boundary is
\begin{equation}
    D_k^\Gamma
    =N_{k-1}^{-1}E_{k-1}^\dagger D_kE_k.
\label{eq:main-coordinate-boundary}
\end{equation}
Its adjoint with respect to the inherited inner products contains the corresponding Gram-matrix factors. Consequently, a direct coordinate implementation requires coherent access to $E_j$, $N_j$, and $N_j^{-1}$ and inherits a dependence on the conditioning of $N_j$. The projected-Dirac route instead uses the ambient sparse boundary access and coherent projections onto $\Gamma$, thereby avoiding an explicit Gram-matrix inversion. This does not make the projectors free; their construction remains part of the input-access assumptions described below.

\section{Quantum Algorithm of GLMY Homology}

In this section we will give the quantum algorithm for computing the Betti numbers of GLMY homology of a digraph $G$. In the quantum algorithm, the Hilbert space is identified with the vector space endowed with inner product discussed in Sec.~\ref{sec:hodge-lapl-oper}. In particular, the Hodge-Laplacian operator $\Delta_k$ or $\Delta_k^{\Gamma}$ are Hermitian operators defined on subspace $\Lambda_k$ or $\Gamma_k$, and the $k$-th Betti number is related to the degeneracy of $\Delta_k^{\Gamma}$ with eigenvalue being $0$.

The digraph $G$ we consider has $n$ vertices labeled by $\{1,\cdots, n\}$, and $d$ ($d\le n-1$) is the maximal path length in $G$. In this case we only need to consider chain groups from $\Gamma_0$ up to $\Gamma_d$. Since $\partial^{\Gamma}_d\Gamma_d=\partial^{\Gamma}_d\Acal_d$, we may replace the highest term $\Gamma_d$ with $\Acal_d$ in our algorithm.

Following the framework of the LGZ algorithm, the quantum procedure has three logically distinct ingredients.

The first ingredient is state preparation. For the $k$-th sector, the state required by the algorithm is the maximally mixed state on the embedded chain space $\Gamma_k$,
\begin{equation}
    u_k\equiv \rho_{\Gamma,k}
    =\frac{\Pi_{\Gamma,k}}{\gamma_k},
    \qquad
    \gamma_k=\dim \Gamma_k,
\label{rho_A_k}
\end{equation}
where $\Pi_{\Gamma,k}$ is the orthogonal projector from $\Lambda_k$ onto $\Gamma_k$. If $\{|g_\ell^{(k)}\rangle\}_{\ell=0}^{\gamma_k-1}$ is any orthonormal basis of $\Gamma_k$, then equivalently
\begin{equation}
    u_k=\frac{1}{\gamma_k}\sum_{\ell=0}^{\gamma_k-1}
    |g_\ell^{(k)}\rangle\langle g_\ell^{(k)}| .
\end{equation}
This basis-independent notation is important because $\Gamma_k=\Acal_k+\partial_{k+1}\Acal_{k+1}$ is an embedded linear subspace, not merely a set of elementary allowed paths.

The second ingredient is operator implementation. We first construct sparse access, or equivalently a block-encoding, for the ambient Dirac operator $B=\partial+\partial^\dagger$ on $\Lambda$. This step uses only the deletion and insertion rules for ordered paths. We then use coherent access to the projector $\Pi$ onto $\Gamma$ to form the projected Dirac operator
\begin{equation}
    B_\Gamma=\Pi B\Pi .
\end{equation}
By Theorem~\ref{thm:total-laplacian}, this operator is the embedded GLMY Dirac operator, and therefore
\begin{equation}
    \Delta^\Gamma=B_\Gamma^2=\Pi B\Pi B\Pi .
\end{equation}
Thus the construction proceeds step by step from the ambient boundary operation to the projected Dirac operator and finally to the embedded Hodge Laplacian.

The third ingredient is phase estimation. Running quantum phase estimation for $\Delta^\Gamma$ on the input state $u_k$ estimates the weight of the zero-eigenvalue subspace. Equivalently, for the purpose of detecting the zero eigenspace only, one may run phase estimation for $B_\Gamma$, since
\begin{equation}
    \ker B_\Gamma=\ker B_\Gamma^2=\ker\Delta^\Gamma .
\end{equation}
Repeating the procedure estimates the normalized Betti number $\beta_k/\dim\Gamma_k$. Multiplication by $\dim\Gamma_k$ gives an estimator of the integer Betti number; exact recovery additionally requires an error below $1/(2\dim\Gamma_k)$ followed by rounding.

\begin{theorem}
\label{thm:phase_estimate}
Let $u_k=\Pi_{\Gamma,k}/\gamma_k$ be the maximally mixed state on $\Gamma_k$, where $\gamma_k=\dim\Gamma_k$. Under an ideal spectral measurement of $\Delta^\Gamma$ (equivalently, of $B_\Gamma$ with the input supported on $\Gamma_k$), the probability assigned to the zero-eigenvalue subspace is
\begin{equation}
    c_k=\frac{\beta_k}{\gamma_k}.
\end{equation}
For a finite-resolution implementation using $B_\Gamma$, define
\begin{equation}
    g_k=\sqrt{\lambda_{\min}^{+}(\Delta_k^\Gamma)},
\end{equation}
where $\lambda_{\min}^{+}$ denotes the smallest positive eigenvalue; if the positive spectrum is empty, all outcomes are zero and no spectral resolution is required. If phase estimation returns an estimate $\widetilde\mu$ whose additive spectral error is smaller than $g_k/4$, classify an outcome as zero when $|\widetilde\mu|<g_k/2$. Except for the phase-estimation failure probability, this rule exactly separates the zero and nonzero spectral components visible from $\Gamma_k$. Repeating the procedure $M$ times gives
\begin{equation}
    \widehat c_k
    =\frac{\#\{i:|\widetilde\mu_i|<g_k/2\}}{M},
    \qquad
    \widehat\beta_k=\gamma_k\widehat c_k.
\end{equation}
The quantity $\widehat\beta_k$ is an estimator rather than an exact identity. If $\gamma_k$ is known and $|\widehat c_k-c_k|<1/(2\gamma_k)$, then the integer Betti number is recovered as
\begin{equation}
    \beta_k=\operatorname{round}(\gamma_k\widehat c_k).
\end{equation}
\end{theorem}
The proof of Theorem~\ref{thm:phase_estimate} is given in Appendix~\ref{proof:phase_estimate}.

\subsection{Encoding of the Paths}
Here we introduce an encoding protocol for paths on directed graphs. Suppose $p_k=v_0\cdots v_k$, where $v_i\in\{1,\ldots,n\}$, is a regular allowed path of length $k$, with $k\le d$. Each vertex register must represent the value $0$ for an absent vertex and the values $1,\ldots,d+1$ for the possible path positions. It therefore contains
\begin{equation}
    l(d)=\left\lceil\log_2(d+2)\right\rceil
\end{equation}
qubits. The path is encoded in an $n l(d)$-qubit state,
\begin{equation}
  \label{eq:1}
  |p_k\rangle
  =\bigotimes_{i=1}^{n}|b^i_1\cdots b^i_{l(d)}\rangle,
\end{equation}
where, if $i=v_{\alpha}$ with $\alpha\in\{0,\ldots,k\}$,
\begin{align}
  \label{eq:4}
  \left(b_1^{v_{\alpha}} b_2^{v_{\alpha}}\cdots b_{l(d)}^{v_{\alpha}}\right)_2
  &=\left(\alpha+1\right)_{10},
\end{align}
and, if $i$ is not a vertex of $p_k$,
\begin{equation}
  \left(b_1^i b_2^i\cdots b_{l(d)}^i\right)_2=\left(0\right)_{10}.
\end{equation}
Here $\left(\cdot\right)_m$ denotes the integer representation in base $m$. For example, using the zero-based vertex labels shown in the second digraph of Fig.~\ref{fig1}(a), the allowed $2$-path $024$ is encoded as $|001\rangle_{0}\otimes|000\rangle_{1}\otimes|010\rangle_{2}\otimes|000\rangle_{3}\otimes|011\rangle_{4}\otimes|000\rangle_{5}$, while the regular $5$-path $320145$ is encoded as $|011\rangle_{0}\otimes|100\rangle_{1}\otimes|010\rangle_{2}\otimes|001\rangle_{3}\otimes|101\rangle_{4}\otimes|110\rangle_{5}$. In this example the maximal path length is $d=5$, so $l(5)=\lceil\log_2 7\rceil=3$.

Although we use this position-label encoding throughout the main construction, the same algorithmic framework can also be implemented with compact path-register encodings analogous to the compact simplex encodings used in streamlined quantum TDA~\cite{mcardle2022streamlinedquantumalgorithmtopological}; this changes only the path representation, not the projected-Dirac and phase-estimation structure.

The encoding above defines the ambient elementary path basis. The preparation of the algorithmic input state $u_k=\Pi_{\Gamma,k}/\gamma_k$ is a separate input-access issue. If a coherent description of an orthonormal basis of $\Gamma_k$ is available, one may prepare a purification of $u_k$ as
\begin{equation}
    |\Phi_k\rangle
    =\frac{1}{\sqrt{\gamma_k}}
    \sum_{\ell=0}^{\gamma_k-1}
    |\ell\rangle_R |\gamma_\ell^{(k)}\rangle_S,
\end{equation}
and then trace out the reference register $R$. This preparation model is the coherent form of direct path-space access.

Under adjacency access supplemented with a coherent reflection about $\Gamma_k$, one instead starts from a purification of the maximally mixed state on the ambient path space. Writing $\lambda_k=\dim\Lambda_k$ and choosing an orthonormal computational basis $\{|p_j\rangle\}_{j=0}^{\lambda_k-1}$ of $\Lambda_k$, prepare
\begin{equation}
    |\Omega_{\Lambda,k}\rangle
    =\frac{1}{\sqrt{\lambda_k}}
      \sum_{j=0}^{\lambda_k-1}|j\rangle_R|p_j\rangle_P.
\end{equation}
Projecting the path register onto $\Gamma_k$ succeeds with probability
\begin{equation}
    \bigl\|(I_R\otimes\Pi_{\Gamma,k})|\Omega_{\Lambda,k}\rangle\bigr\|^2
    =\frac{\gamma_k}{\lambda_k}
    =\zeta_k.
\end{equation}
After amplitude amplification, the normalized projected state is a purification of $u_k$, because tracing out $R$ gives $\Pi_{\Gamma,k}/\gamma_k$. This route has the characteristic cost $\mathcal O(\zeta_k^{-1/2})$. A pure uniform superposition of paths would not produce the required maximally mixed state after projection; the reference register in $|\Omega_{\Lambda,k}\rangle$ is essential. State preparation is therefore an explicit resource rather than an automatic consequence of adjacency-list input.

In the stronger direct path-space input model, the input provides more direct access to $\Gamma_k$, for example through an explicit path-space description, a path list, a succinct path-space specification, or a coherent loading routine. In that case the preparation cost is the cost of this loading operation rather than a Grover search over all of $\Lambda_k$. This distinction will be used in the complexity analysis below.

\subsection{Encoding and sparse access of the boundary operator}
\label{subsec:boundary-sparse-access}

The path encoding introduced above makes the GLMY boundary rule particularly transparent.  For an elementary path $p_k=v_0\cdots v_k$, the boundary operator deletes one path position at a time,
\begin{equation}
    \partial_k|p_k\rangle
    =\sum_{r=0}^{k}(-1)^r
    |v_0\cdots \widehat v_r\cdots v_k\rangle .
\label{eq:main-boundary-action}
\end{equation}
In the position-label representation, deleting the $r$-th path position means erasing the vertex carrying label $r+1$ and decreasing every later position label by one.  The adjoint operation reverses this update: it inserts an absent vertex at a chosen position and shifts the later labels.  Both operations consist only of reversible comparisons, controlled increments or decrements, and a phase $(-1)^r$ determined by the deleted or inserted position.

These deletion and insertion moves provide sparse access to the ambient path Dirac operator
\begin{equation}
    B=\partial+\partial^\dagger
\end{equation}
on $\Lambda=\bigoplus_{k=0}^{d}\Lambda_k$.  A path of length at most $d$ has at most $d+1$ deletion neighbours and at most $n(d+1)$ insertion candidates.  Hence the row sparsity of $B$ obeys
\begin{equation}
    s_B=O(nd),
\label{eq:main-dirac-sparsity}
\end{equation}
where $n$ is the number of vertices and $d$ is the maximal path length considered.  Standard sparse-matrix techniques therefore convert the reversible location and value oracles into a block-encoding $U_B$ satisfying
\begin{equation}
    (\langle 0^{a_B}|\otimes I)U_B(|0^{a_B}\rangle\otimes I)
    =\frac{B}{\alpha_B},
\label{eq:main-B-block-encoding}
\end{equation}
where $a_B$ is the number of ancillary qubits and $\alpha_B\geq\|B\|$ is the normalization factor.  For the direct sparse construction with matrix entries in $\{0,\pm1\}$, one may take $\alpha_B=O(s_B)=O(nd)$.  The explicit reversible deletion circuit, the corresponding insertion circuit, and a direct block-encoding construction are given in Appendix~\ref{app:boundary-circuit}.

\subsection{Projected-Laplacian simulation under direct path-space access}
\label{subsec:projected-laplacian-simulation}

The operator $B$ acts on the full regular-path space $\Lambda$, whereas the GLMY chain complex is the embedded subspace
\begin{equation}
    \Gamma=\bigoplus_{k=0}^{d}\Gamma_k\subseteq\Lambda.
\end{equation}

The additional task is therefore to keep the evolution inside $\Gamma$. In the direct path-space access model, this is achieved by coherent loaders for all degree sectors. For each $j\le d$, let $\gamma_j=\dim\Gamma_j$ and let $\{|\gamma^{(j)}_\ell\rangle\}_{\ell=0}^{\gamma_j-1}$ be an orthonormal basis of $\Gamma_j$. We assume that a unitary $V_{\Gamma,j}$ and its inverse can be implemented such that
\begin{equation}
    V_{\Gamma,j}|\ell\rangle_P
    =|\gamma^{(j)}_\ell\rangle_P,
    \qquad 0\leq\ell<\gamma_j .
\label{eq:main-Gamma-loader}
\end{equation}
Applying $V_{\Gamma,j}^\dagger$, checking whether the recovered label lies in the valid range, and applying $V_{\Gamma,j}$ again implements $\Pi_{\Gamma,j}$. A degree-controlled direct sum of these routines implements
\begin{equation}
    \Pi=\bigoplus_{j=0}^{d}\Pi_{\Gamma,j},
\end{equation}
the projector needed in the total operator $B_\Gamma=\Pi B\Pi$. The sector loader $V_{\Gamma,k}$ prepares the input state, whereas the controlled family $\{V_{\Gamma,j}\}_{j=0}^{d}$ implements the projectors used during phase estimation.

Combining this projector with the ambient Dirac operator gives
\begin{equation}
    B_\Gamma=\Pi B\Pi .
\label{eq:main-projected-dirac}
\end{equation}
The meaning is simple: first retain the component in $\Gamma$, next apply one ambient deletion or insertion step, and finally discard every component that has left $\Gamma$.  On $\Gamma$, Theorem~\ref{thm:total-laplacian} identifies $B_\Gamma$ with the Dirac operator of the embedded GLMY chain complex,
\begin{equation}
    B_\Gamma=\partial^\Gamma+(\partial^\Gamma)^{*} .
\end{equation}
Consequently,
\begin{equation}
    \Delta^\Gamma
    =B_\Gamma^2
    =\Pi B\Pi B\Pi
    =(\partial^\Gamma)^{*}\circ\partial^\Gamma
     +\partial^\Gamma\circ(\partial^\Gamma)^{*}
\label{eq:main-embedded-laplacian}
\end{equation}
is exactly the embedded Hodge Laplacian, rather than merely the ambient Laplacian projected at its two ends.

At the circuit level, the three selected blocks are composed in the same order as the operator product: a projector block-encoding, the ambient Dirac block-encoding $U_B$, and a second projector block-encoding give a unitary whose selected block is $B_\Gamma/\alpha_B$.  Repeating this product construction with independent ancillary registers gives a unitary $U_\Delta$ satisfying
\begin{equation}
    (\langle0^{a_\Delta}|\otimes I)U_\Delta
    (|0^{a_\Delta}\rangle\otimes I)
    =\frac{\Delta^\Gamma}{\alpha_B^2}.
\end{equation}
Thus the nonunitary Laplacian is not applied directly as a gate; instead, its normalized matrix is stored in a selected block of the larger unitary $U_\Delta$.

To simulate the Hamiltonian evolution, set
\begin{equation}
    H_\Delta=\frac{\Delta^\Gamma}{\alpha_\Delta},
    \qquad \alpha_\Delta=\alpha_B^2,
    \qquad \|H_\Delta\|\leq1.
\end{equation}
The simulation follows the same order as the quantum circuit.  First, because the block-encoding $U_\Delta$ need not itself be Hermitian, one adds a single control qubit and combines $U_\Delta$ with $U_\Delta^\dagger$ to obtain a Hermitian unitary $\mathcal U_\Delta$ whose selected block is still $H_\Delta$.  Second, reflecting about the selected ancilla state and multiplying this reflection by $\mathcal U_\Delta$ produces a qubitized walk $W_\Delta$.  For an eigenpair
\begin{equation}
    H_\Delta|\psi_j\rangle=x_j|\psi_j\rangle,
    \qquad x_j=\frac{\nu_j}{\alpha_\Delta},
\end{equation}
where $\nu_j$ is the corresponding eigenvalue of $\Delta^\Gamma$, the walk acts in a two-dimensional invariant subspace as a rotation through an angle $\theta_j$ satisfying
\begin{equation}
    \cos\theta_j=x_j.
\end{equation}
Thus the unknown eigenvalue is encoded coherently as a walk rotation angle; it is not estimated or supplied to the circuit as a classical number.

Quantum signal processing is then applied to this walk.  A classical phase-synthesis procedure uses only $\alpha_\Delta t$ and the target simulation error to choose fixed phases $\phi_0,\ldots,\phi_m$ such that the corresponding polynomial $P_m(x)$ uniformly approximates $e^{-i\alpha_\Delta tx}$ on $[-1,1]$.  The quantum circuit alternates calls to $W_\Delta$ with rotations about the signal reflection using these phases.  Consequently, on every eigencomponent, the same circuit produces
\begin{equation}
    P_m(x_j)\approx e^{-i\alpha_\Delta t x_j}
    =e^{-i\nu_jt}.
\end{equation}
By linearity, this realizes $e^{-i\Delta^\Gamma t}$ on an arbitrary superposition.  This construction is commonly described as qubitization followed by quantum signal processing, or equivalently as the Hermitian-matrix instance of quantum singular value transformation.  No individual eigenvalue is required in advance, and the block-encoding ancillas are used coherently rather than measured and postselected after each query~\cite{PhysRevLett.118.010501,Low2019hamiltonian,10.1145/3313276.3316366}.

For Betti-number estimation it is often more economical to apply phase estimation directly to $B_\Gamma$, because
\begin{equation}
    \ker B_\Gamma=\ker\Delta^\Gamma,
\end{equation}
and this avoids the normalization overhead caused by explicitly squaring the block-encoded operator.  Appendix~\ref{app:projected-simulation} gives the projector circuit, the product block-encodings, the qubitization step, and the subsequent QSVT Hamiltonian simulation in the same order as they are used by the algorithm.

\subsection{Complexity Analysis}
\label{subsec:complexity-main}

We state the complexity for estimating the normalized Betti number
\begin{equation}
    c_k=\frac{\beta_k}{\gamma_k},
    \qquad \gamma_k=\dim\Gamma_k,
\end{equation}
to additive error $\varepsilon$ with failure probability at most $\delta$.  Since one phase-estimation run returns the zero-eigenvalue outcome with probability $c_k$, standard sampling requires
\begin{equation}
    M=O\!\left(\frac{\log(1/\delta)}{\varepsilon^2}\right)
\label{eq:main-number-samples}
\end{equation}
independent repetitions.

To make the gate bounds explicit, define
\begin{equation}
    l(d)=\left\lceil\log_2(d+2)\right\rceil,
    \qquad
    q_P=n l(d),
\end{equation}
where $q_P$ is the number of qubits in the position-label path register.  As in the standard query-complexity model, one coherent query to the directed-edge oracle is counted as unit cost $O(1)$.  Scanning and reversibly updating the $n$ position labels and checking at most $O(d)$ directed edges therefore gives the following cost for one use of the ambient Dirac block-encoding:
\begin{equation}
    T_B=O\!\left(nl(d)+d\right),
\label{eq:main-explicit-TB}
\end{equation}
up to the polylogarithmic arithmetic and precision factors suppressed below by $\widetilde O$.  Its row sparsity and block-encoding normalization obey
\begin{equation}
    s_B=O(nd),
    \qquad
    \alpha_B=O(nd).
\label{eq:main-explicit-alphaB}
\end{equation}
Define
\begin{equation}
    g_k=\sqrt{\lambda_{\min}^{+}(\Delta_k^\Gamma)},
\end{equation}
when the positive spectrum is nonempty; if it is empty, the $k$-sector is entirely harmonic and the spectral-resolution step is trivial. The operator used in phase estimation is the total projected Dirac operator $B_\Gamma=\Pi B\Pi$. For an input supported on $\Gamma_k$, the nonzero Dirac eigenvalue magnitudes that carry spectral weight are the square roots of the positive eigenvalues of $\Delta_k^\Gamma$, and hence are at least $g_k$. The normalized spectral separation is $g_k/\alpha_B$, so distinguishing zero from the nonzero spectrum requires
\begin{equation}
    \kappa_k=\frac{\alpha_B}{g_k}
    =O\!\left(\frac{nd}{g_k}\right)
\label{eq:main-kappa-definition}
\end{equation}
block-encoding uses, up to logarithmic simulation and phase-estimation factors. Direct phase estimation of $B_\Gamma$ avoids explicitly forming the squared block-encoding, which would square both the normalization and the relevant spectral scale.

Under adjacency access supplemented with a coherent $\Gamma$-reflection oracle (denoted standard access below), define
\begin{equation}
    \zeta_k=\frac{\dim\Gamma_k}{\dim\Lambda_k}.
\end{equation}
For an explicit bound, we make the input-model promise that coherent reflections $R_{\Gamma,j}=2\Pi_{\Gamma,j}-I$ are supplied for every degree sector, with uniform reversible cost
\begin{equation}
    C_{\Pi}^{\mathrm{std}}
    =O\!\left(nl(d)+d\right).
\label{eq:main-standard-projector-promise}
\end{equation}
The particular reflection $R_{\Gamma,k}$ and its controlled marking operation are used for amplitude amplification, while the degree-controlled direct sum of the family $\{R_{\Gamma,j}\}$ implements the total projector $\Pi$ inside $B_\Gamma$. This is a genuine access assumption: an ordinary adjacency oracle alone does not in general provide these reflections. Under this promise, amplitude amplification prepares the input state with cost
\begin{equation}
    \widetilde O\!\left[
       \left(nl(d)+d\right)\zeta_k^{-1/2}
    \right],
\end{equation}
and one query to the block-encoding of the total operator $B_\Gamma=\Pi B\Pi$ also costs $O(nl(d)+d)$ up to constants, provided the degree-controlled reflection about $\Gamma=\bigoplus_j\Gamma_j$ is supplied with this cost. Therefore
\begin{equation}
    T_{\mathrm{tot}}^{\mathrm{std}}
    =\widetilde O\!\left[
      \left(nl(d)+d\right)
      \left(
         \zeta_k^{-1/2}+\frac{nd}{g_k}
      \right)
      \frac{\log(1/\delta)}{\varepsilon^2}
    \right].
\label{eq:main-complexity-standard}
\end{equation}
The factor $\zeta_k^{-1/2}$ is the state-preparation bottleneck and can be exponentially large when $\Gamma_k$ is sparse inside $\Lambda_k$.

Under direct path-space access, the input supplies $V_{\Gamma,j}$ and $V_{\Gamma,j}^{\dagger}$ for all degree sectors. In the parallel coherent-loading model, define
\begin{equation}
    L_j=O\!\left(\log\!\left[\gamma_j n l(d)\right]\right),
    \qquad
    L_\Pi=\max_{0\le j\le d}L_j
    =O\!\left(\log\!\left[\gamma_{\max} n l(d)\right]\right),
\label{eq:main-explicit-loader}
\end{equation}
where $\gamma_{\max}=\max_j\gamma_j$. The input purification in the $k$-th sector costs $O(L_k)$, while the degree-controlled projector in each query to $B_\Gamma$ costs $O(L_\Pi)$. Hence one projected-Dirac query costs
\begin{equation}
    O\!\left(nl(d)+d+L_\Pi\right).
\end{equation}
The direct-access bound is therefore
\begin{equation}
\begin{aligned}
    T_{\mathrm{tot}}^{\mathrm{dir}}
    =\widetilde O\!\Bigg[
      \Bigg(&
        \log\!\left[\gamma_k n l(d)\right]
        +\left(
        nl(d)+d
        +\log\!\left[\gamma_{\max} n l(d)\right]
        \right)\frac{nd}{g_k}
      \Bigg)
      \frac{\log(1/\delta)}{\varepsilon^2}
    \Bigg].
\end{aligned}
\label{eq:main-complexity-direct}
\end{equation}
The quantities $L_j$ and $L_\Pi$ are circuit-depth assumptions for a parallel data-loading architecture. The formulas above should therefore be read as query and circuit-depth bounds. They do not include the total gate count, classical memory, or parallel hardware required to store and address the path data, which may scale with the size of that data structure.

Here $\widetilde O$ hides only polylogarithmic factors associated with reversible arithmetic, Hamiltonian-simulation accuracy, and phase-estimation accuracy; it does not hide the displayed sampling factor $\varepsilon^{-2}$.  Under standard access, no unconditional exponential speedup is claimed because $\zeta_k^{-1/2}$ may be exponential and because Eq.~\eqref{eq:main-standard-projector-promise} is an additional oracle promise.  Under direct access, the logarithmic loader depth in Eq.~\eqref{eq:main-explicit-loader} is likewise a stronger data-access assumption.  Subject to these assumptions and polynomially bounded $g_k^{-1}$, the quantum runtime is polynomial in the path-space description and may be exponentially smaller than classical path-homology algorithms when the relevant path length grows.  
These qualifications are consistent with recent resource-aware and complexity-theoretic analyses of quantum TDA, which show that a claimed quantum advantage must be assessed together with the input-access model, state-preparation cost, spectral-resolution requirement, output precision, and the computational hardness of the underlying homology problem~\cite{PRXQuantum.5.010319,PRXQuantum.4.040349,Crichigno2024}. In the present GLMY setting, these considerations are reflected explicitly in the distinction between standard digraph access and direct path-space access.
The algorithm efficiently estimates $c_k$. Recovering the exact integer additionally requires knowing $\gamma_k$ and estimating $c_k$ to error smaller than $1/(2\gamma_k)$, after which one sets $\beta_k=\operatorname{round}(\gamma_k\widehat c_k)$.

The derivation of Eqs.~\eqref{eq:main-complexity-standard} and \eqref{eq:main-complexity-direct}, including the state-preparation, projected-Dirac, spectral-resolution, and sampling factors, is given in Appendix~\ref{app:detailed-complexity}.

\section{Conclusion and Outlook}

In this work, we developed a quantum algorithmic framework for computing the GLMY homology of digraphs. The GLMY setting differs from the simplicial setting in two important ways: chains are built from ordered paths, and the embedded chain space $\Gamma_k$ need not be represented by a canonical orthonormal elementary-path basis. These features prevent a direct substitution of simplices by paths in existing quantum TDA algorithms.

Our main contribution is to show how the LGZ-type strategy can nevertheless be adapted to directed path homology. First, we introduced a quantum encoding of ordered paths and gave an explicit deletion and insertion description of the corresponding boundary and adjoint-boundary operations. Second, we proved the projected-Dirac identity $\Delta^\Gamma=\Pi B\Pi B\Pi$, which avoids the direct construction of the embedded Hodge Laplacian through Gram matrices and their inverses. Third, we separated the resources needed for the algorithm into state preparation, sparse or block-encoded access to the path Dirac operator, spectral resolution, and sampling precision. This separation clarifies that the attainable quantum advantage depends on the input model rather than being unconditional.

The main limitation is state preparation. Under standard digraph access, preparing a state supported on $\Gamma_k$ may require amplitude amplification with cost $\zeta_k^{-1/2}$, which can be large when the embedded path space is sparse. Under direct path-space access, this bottleneck is replaced by the cost of coherently loading the path subspace, but that is a stronger assumption. The spectral-resolution parameter $\kappa_k$ and the precision required to recover an integer Betti number may also be significant in difficult instances. Thus the most appropriate interpretation of the result is an input-model-dependent quantum framework for normalized GLMY Betti-number estimation.

Several application domains appear particularly promising for the proposed framework. (i)~\textit{Temporal networks:} Path homology has already been applied to temporal networks~\cite{chowdhury2022path}, where events are naturally time-ordered and the path space is constrained by temporal causality, leading to moderate $|\Gamma_k|$ and making the direct path-space input assumption realistic. Examples include email communication networks, financial transaction graphs, and biological signaling cascades. (ii)~\textit{Deep neural network analysis:} GLMY homology has been used to study topological properties of deep feedback networks~\cite{chowdhury2019path}. Neural architectures with directed information flow (e.g., feedforward layers, residual connections) naturally admit a digraph description, and their structured connectivity may yield favorable $\zeta_k$ values. (iii)~\textit{Molecular and materials science:} Applications of path homology to molecular systems~\cite{chen2023path} benefit from the fact that many chemical bonds (coordination bonds, hydrogen bonds) are directional. For small-to-medium molecular datasets where $|\Gamma_k|$ remains polynomially bounded, the quantum algorithm may offer practical advantages. (iv)~\textit{Quiver representations:} The generalization of GLMY homology to quivers~\cite{GrigoryanMuranovVershininYau+2018+1319+1337} suggests potential applications in representation theory and algebraic geometry, where computing homological invariants of quivers is a central computational task.

There are several directions for future work.

(i)~\textit{State preparation and path encoding.} The dominant computational bottleneck is the preparation of superpositions over $\Gamma_k$, which requires $\mathcal{O}(\zeta_k^{-1/2})$ Grover iterations under standard digraph input. More efficient state-preparation protocols and specialized quantum membership oracles for $\Gamma_k$---particularly for structured families such as temporal networks and directed acyclic graphs, where causality constrains admissible paths~\cite{Ameneyro2022}---could substantially reduce this overhead. In parallel, a compact path encoding \`a la McArdle et al.~\cite{mcardle2022streamlinedquantumalgorithmtopological} that stores the ordered vertex tuple $(v_0,\dots,v_k)$ using only $\mathcal{O}(k\log n)$ qubits, rather than the current $\mathcal{O}(n\log d)$ vertex-position encoding, would yield an exponential qubit reduction when $k\ll n$. The main challenges are implementing the boundary operator on list-encoded paths and handling the resulting non-orthogonality of path states.

(ii)~\textit{Persistent GLMY homology and generalizations.} The persistent Dirac operator framework of Ameneyro et al.~\cite{Ameneyro2024} extends naturally to the GLMY setting. For a filtered digraph $G_{\varepsilon}$, the persistent path space $\Gamma_k(\varepsilon)=\Acal_k(\varepsilon)+\partial_{k+1}(\varepsilon)\Acal_{k+1}(\varepsilon)$ and the persistent Laplacian $\Delta(\varepsilon)$ satisfy Theorems~\ref{thm:total-laplacian} and~\ref{thm:phase_estimate} level-wise, enabling Betti number estimation via phase estimation at each filtration level without explicit dual-map construction. The principal technical challenges are: uniformly bounding the spectral gap $g_k(\varepsilon)$ across all levels, controlling the multiplicative accumulation of state-preparation cost over the filtration scan, and handling the non-trivial inclusion maps $\Gamma_k(\varepsilon)\hookrightarrow\Gamma_k(\varepsilon')$ which involve the condition $\partial x\in\Acal_{k-1}$ rather than simple vertex insertions. Beyond persistence, the framework naturally generalizes to weighted digraphs~\cite{lin2019weighted}, quivers~\cite{GrigoryanMuranovVershininYau+2018+1319+1337}, and weighted hypergraphs~\cite{muranov2023homology}.

(iii)~\textit{GLMY Harmonic Persistence.} Gyurik et al.~\cite{gvys-hl8h} introduced the Harmonic Persistence problem for simplicial complexes---determining whether a harmonic cycle persists across a filtration---and proved it $\BQP_1$-hard and contained in $\BQP$, providing complexity-theoretic evidence for exponential quantum advantage. Our Hodge decomposition $\Gamma_k=\operatorname{Ker}\Delta^{\Gamma}_k\oplus\partial^{\Gamma}_{k+1}\Gamma_{k+1}\oplus(\partial^{\Gamma}_k)^*\Gamma_{k-1}$ and Theorem~\ref{thm:total-laplacian} provide the algebraic foundation for a GLMY analogue: given a harmonic path $h\in\operatorname{Ker}\Delta^{\Gamma}_k(\varepsilon_0)$, determine whether it remains harmonic at later filtration levels. Key obstacles include deterministic preparation of harmonic guiding states and tracking harmonicity through $\partial x\in\Acal_{k-1}$-constrained inclusion maps.

(iv)~\textit{NISQ-friendly implementation.} Akhalwaya et al.~\cite{ICLR2024_9675b27e} replaced quantum phase estimation with stochastic rank estimation (SRE) for simplicial homology, dramatically reducing circuit depth and enabling execution on noisy quantum processors. Adapting SRE to GLMY homology faces the obstacle that a natural generating set of $\Gamma_k$ need not be orthonormal. Three possible strategies are: (a) use the ambient orthonormal register together with a block-encoding of the correct embedded operator $\Delta^\Gamma=(\Pi B\Pi)^2$ and sample a purification of $u_k$; (b) incorporate the induced Gram matrix into a coordinate-based rank estimator, at the cost of its construction and conditioning; or (c) apply polynomial or stochastic spectral filtering directly to the total projected Dirac operator $B_\Gamma=\Pi B\Pi$, using a finite zero window controlled by $g_k$. Merely replacing $\Delta_k^\Gamma$ by the endpoint projection $\Pi_{\Gamma,k}\Delta_k\Pi_{\Gamma,k}$ is not valid in general, because the intermediate projection in $\Pi B\Pi B\Pi$ is essential.

\section*{Acknowledgments}
This work is supported by Science Challenge Project (Grant No. TZ2025017), National Key Research and Development Program of China (Grants No. 2021YFA1402104 and No. 2021YFA0718302), National Natural Science Foundation of China (Grant No. 12075310) and Natural Science Foundation of Shandong Province (Grant No. ZR2025QC1476). The authors declare no competing interests.

\section*{Data availability statement}
No new data were generated or analysed during this study.

\appendix
\renewcommand{\theHequation}{appendix.\theequation}
\section{Proof of Theorem~\ref{thm:dual-commute}}
\label{proof:dual-commute}
\begin{proof}[Proof of Theorem~\ref{thm:dual-commute}]
       The proof is a direct computation using the defining relations of the adjoint operators and the fact that the embeddings $\iota_k$, $\iota_{k-1}$ are isometric with respect to the inherited inner products.
    
    Take arbitrary $x\in\Gamma_k$ and $y\in\Gamma_{k-1}$. We compare the inner products of both sides of the claimed equality with $x$ in $\Lambda_k$:
    \begin{align}
        &\langle (\partial_k^{\Gamma})^{*}(y),\, x\rangle_{\Gamma}\\
        &= \langle y,\, \partial_k^{\Gamma}(x)\rangle_{\Gamma}
        &&\text{(definition of $(\partial_k^{\Gamma})^{*}$)}\nonumber\\
        &= \langle \iota_{k-1}(y),\, \iota_{k-1}\circ\partial_k^{\Gamma}(x)\rangle_{\Lambda}
        &&\text{(since $\iota_{k-1}$ is isometric)}\nonumber\\
        &= \langle \iota_{k-1}(y),\, \partial_k\circ x\rangle_{\Lambda}
        &&\text{(by Eq.~\eqref{Eq:partGam-Part})}\nonumber\\
        &= \langle \partial_k^{*}\circ \iota_{k-1}(y),\, x\rangle_{\Lambda}
        &&\text{(definition of $\partial_k^{*}$)}\nonumber\\
        &= \langle \pi_k\circ\partial_k^{*}\circ \iota_{k-1}(y),\, x\rangle_{\Gamma}
        &&\text{(since $x\in\Gamma_k$, the orthogonal projection onto $\Gamma_k$ does not}\nonumber\\
        &&&\text{change the inner product).}\nonumber
    \end{align}
    Both $(\partial_k^{\Gamma})^{*}(y)$ and $\pi_k\circ\partial_k^{*}\circ \iota_{k-1}(y)$ lie in $\Gamma_k$, and we have shown that their difference is orthogonal to every element $x$ of $\Gamma_k$. Hence the difference vanishes, and the equality
    \begin{equation}
        (\partial_k^{\Gamma})^{*}=\pi_k\circ\partial_k^{*}\circ \iota_{k-1}
    \end{equation}
    holds for all $y\in\Gamma_{k-1}$. 
\end{proof}

Combining the equation $\iota_{k-1}\circ\partial_k^{\Gamma}=\partial_k\circ \iota_k$ with Theorem~\ref{thm:dual-commute} we obtain the following commutative diagram (where each slanted bottom arrow implicitly includes the projection onto the corresponding $\Gamma$-space):
\begin{equation}\label{apeq:full-commute-diag}
    \xymatrix{
        \ar@/^/[rr]&&\Gamma_{k+1}\ar@/^/[rr]^{\partial_{k+1}^{\Gamma}}\ar[ll]\ar[dd]^{\iota_{k+1}}&&\Gamma_k\ar[ll]^{(\partial_{k+1}^{\Gamma})^{*}}\ar@/^/[rr]^{\partial_{k}^{\Gamma}}\ar[dd]^{\iota_k}&&\Gamma_{k-1}\ar[ll]^{(\partial_{k}^{\Gamma})^{*}}\ar@/^/[rr]^{\partial_{k-1}^{\Gamma}}\ar[dd]^{\iota_{k-1}}&&\ar[ll]\\
        &&&&&&&&\\
        \ar@/^/[rr]&&\Lambda_{k+1}\ar@/^/[uu]^{\pi_{k+1}}\ar@/^/[rr]^{\partial_{k+1}}\ar[ll]&&\Lambda_{k}\ar@/^/[uu]^{\pi_{k}}\ar[ll]^{\partial_{k+1}^{*}}\ar@/^/[rr]^{\partial_k}&&\Lambda_{k-1}\ar@/^/[uu]^{\pi_{k-1}}\ar[ll]^{\partial_{k}^{*}}\ar@/^/[rr]^{\partial_{k-1}}&&\ar[ll]
    }
\end{equation}

\section{Proof of Theorem~\ref{thm:total-laplacian}}
\label{proof:total-laplacian}

\begin{proof}
Let $\Pi=\bigoplus_k\pi_k$ be the orthogonal projection from $\Lambda=\bigoplus_k\Lambda_k$ onto $\Gamma=\bigoplus_k\Gamma_k$, and let $B=\partial+\partial^\dagger$ be the ambient path Dirac operator on $\Lambda$. We first show that the projected Dirac operator $B_\Gamma=\Pi B\Pi$, restricted to $\Gamma$, is the Dirac operator of the embedded GLMY chain complex.

Take $x\in\Gamma_k$. Since $\Pi x=x$, we have
\begin{equation}
    B_\Gamma x=\Pi Bx
    =\pi_{k-1} \circ\partial_k x+\pi_{k+1}\circ\partial_{k+1}^*x .
\end{equation}
The first term is the embedded boundary term. Indeed, by the defining relation
\begin{equation}
    \iota_{k-1}\circ\partial_k^\Gamma=\partial_k\circ\iota_k,
\end{equation}
we have $\partial_k x=\partial_k^\Gamma x\in\Gamma_{k-1}$ after identifying $\Gamma_{k-1}$ with its image in $\Lambda_{k-1}$. Hence
\begin{equation}
    \pi_{k-1}\circ\partial_k x=\partial_k^\Gamma x .
\end{equation}
For the second term, Theorem~\ref{thm:dual-commute}, with the index shifted by one, gives
\begin{equation}
    (\partial_{k+1}^\Gamma)^*=
    \pi_{k+1}\circ\partial_{k+1}^*\circ\iota_k .
\end{equation}
Since $x\in\Gamma_k$, this implies
\begin{equation}
    \pi_{k+1}\circ\partial_{k+1}^*x=(\partial_{k+1}^\Gamma)^*x .
\end{equation}
Therefore
\begin{equation}
    B_\Gamma x
    =\partial_k^\Gamma x+(\partial_{k+1}^\Gamma)^*x .
\label{eq:BGamma-action-appendix}
\end{equation}
As $x$ was arbitrary, this proves that on the total embedded chain space
\begin{equation}
    B_\Gamma=\partial^\Gamma+(\partial^\Gamma)^* .
\end{equation}

Squaring this identity gives
\begin{align}
    B_\Gamma^2
    &=\bigl(\partial^\Gamma+(\partial^\Gamma)^*\bigr)^2\nonumber\\
    &=(\partial^\Gamma)^*\circ\partial^\Gamma
      +\partial^\Gamma\circ(\partial^\Gamma)^* .
\end{align}
The two remaining cross terms vanish because $(\partial^\Gamma)^2=0$ and $((\partial^\Gamma)^*)^2=0$. The right-hand side is precisely the embedded Hodge Laplacian $\Delta^\Gamma$. Thus
\begin{equation}
    \Delta^\Gamma=B_\Gamma^2=(\Pi B\Pi)^2=\Pi B\Pi B\Pi,
\end{equation}
where the last equality uses that the operator is restricted to the range of $\Pi$. This proves the theorem.
\end{proof}

\section{Proof of Theorem~\ref{thm:phase_estimate}}
\label{proof:phase_estimate}

\begin{proof}
Let $E^{(k)}_0$ be the spectral projector of $\Delta^\Gamma_k$ onto its zero-eigenvalue subspace. An ideal spectral measurement samples the spectral measure of $\Delta^\Gamma_k$ with respect to
\begin{equation}
    u_k=\frac{\Pi_{\Gamma,k}}{\gamma_k},
    \qquad \gamma_k=\dim\Gamma_k.
\end{equation}
Therefore
\begin{align}
    p_0
    &=\tr\bigl(E^{(k)}_0u_k\bigr)\nonumber\\
    &=\frac{1}{\gamma_k}\tr\bigl(E^{(k)}_0\Pi_{\Gamma,k}\bigr)
      =\frac{\rk E^{(k)}_0}{\gamma_k}
      =\frac{\beta_k}{\gamma_k}.
\end{align}
The same probability is obtained by applying the spectral measurement to the total projected Dirac operator $B_\Gamma$. Indeed, $B_\Gamma^2=\Delta^\Gamma$, so $\ker B_\Gamma=\ker\Delta^\Gamma$, and the input $u_k$ selects the $k$-th graded component of this kernel.

For finite-resolution phase estimation on $B_\Gamma$, set
\begin{equation}
    g_k=\sqrt{\lambda_{\min}^{+}(\Delta_k^\Gamma)}.
\end{equation}
If the phase-estimation error is smaller than $g_k/4$, a returned estimate $\widetilde\mu$ is classified as zero whenever $|\widetilde\mu|<g_k/2$. This classification is correct except on the phase-estimation failure event. Thus
\begin{equation}
    \widehat c_k
    =\frac{\#\{i:|\widetilde\mu_i|<g_k/2\}}{M}
\end{equation}
is an estimator of $c_k=\beta_k/\gamma_k$, and
\begin{equation}
    \widehat\beta_k=\gamma_k\widehat c_k
\end{equation}
is the corresponding estimator of the Betti number. If $|\widehat c_k-c_k|<1/(2\gamma_k)$, then exact integer recovery is obtained by
\begin{equation}
    \beta_k=\operatorname{round}(\gamma_k\widehat c_k).
\end{equation}
\end{proof}

\section{Circuit construction and detailed complexity analysis}
\label{app:detailed-complexity}

This appendix gives the circuit-level construction that is summarized in Secs.~\ref{subsec:boundary-sparse-access} and \ref{subsec:projected-laplacian-simulation}, followed by a derivation of the complexity formulas stated in Sec.~\ref{subsec:complexity-main}.

\subsection{Basic parameters and state preparation}

Let $G=(V,E)$ be a digraph with $|V|=n$, let $d$ be the maximal path length considered, and define

\begin{equation}
    l(d)=\left\lceil\log_2(d+2)\right\rceil,
    \qquad
    q_P=nl(d).
\label{eq:app-path-register-size}
\end{equation}
For each $k\leq d$, let
\begin{equation}
    \lambda_k=\dim\Lambda_k,
    \qquad
    \gamma_k=\dim\Gamma_k,
    \qquad
    \zeta_k=\frac{\gamma_k}{\lambda_k}.
\end{equation}
The input to the $k$-th sector is the maximally mixed state
\begin{equation}
    u_k=\frac{\Pi_{\Gamma,k}}{\gamma_k}.
\label{eq:app-maximally-mixed}
\end{equation}
If $\{|\gamma^{(k)}_\ell\rangle\}_{\ell=0}^{\gamma_k-1}$ is an orthonormal basis of $\Gamma_k$, a purification is
\begin{equation}
    |\Phi_k\rangle
    =\frac{1}{\sqrt{\gamma_k}}
      \sum_{\ell=0}^{\gamma_k-1}
      |\ell\rangle_R|\gamma^{(k)}_\ell\rangle_P.
\label{eq:app-purification}
\end{equation}
Tracing out the reference register $R$ gives Eq.~\eqref{eq:app-maximally-mixed}.

For adjacency access supplemented with coherent embedded-subspace reflections (denoted standard access below), begin with a purification of the maximally mixed state on the ambient path space,
\begin{equation}
    |\Omega_{\Lambda,k}\rangle
    =\frac{1}{\sqrt{\lambda_k}}
      \sum_{j=0}^{\lambda_k-1}|j\rangle_R|p_j\rangle_P,
\end{equation}
where $\{|p_j\rangle\}$ is an orthonormal computational basis of $\Lambda_k$.  Projecting the path register onto $\Gamma_k$ succeeds with probability
\begin{equation}
    \bigl\|(I_R\otimes\Pi_{\Gamma,k})|\Omega_{\Lambda,k}\rangle\bigr\|^2
    =\frac{\tr\Pi_{\Gamma,k}}{\lambda_k}
    =\zeta_k.
\end{equation}
We assume that the standard-access input supplies coherent reflections
\begin{equation}
    R_{\Gamma,j}=2\Pi_{\Gamma,j}-I,
    \qquad 0\le j\le d,
\end{equation}
or equivalent controlled marking operations, with a uniform reversible cost
\begin{equation}
    C_{\mathrm{std}}
    =O\!\left(nl(d)+d\right).
\label{eq:app-standard-reflection-cost}
\end{equation}
The $j=k$ reflection is used for input-state preparation, while the degree-controlled family implements the total projector in $B_\Gamma$. This is an explicit input-model promise: a usual adjacency oracle does not by itself implement reflections about these general linear subspaces. Amplitude amplification then uses $O(\zeta_k^{-1/2})$ coherent reflection calls and gives
\begin{equation}
    T_{\mathrm{prep}}^{\mathrm{std}}
    =\widetilde O\!\left[
      \left(nl(d)+d\right)\zeta_k^{-1/2}
    \right].
\label{eq:app-prep-standard}
\end{equation}

In the direct path-space access model, the input supplies a coherent basis-change unitary and its inverse for every degree sector,
\begin{equation}
    V_{\Gamma,j}|\ell\rangle_P
    =|\gamma^{(j)}_\ell\rangle_P,
    \qquad 0\leq\ell<\gamma_j,
\label{eq:app-loader}
\end{equation}
where the states $|\ell\rangle_P$ are computational-basis reference labels. Under the assumed parallel coherent-loading data structure, define
\begin{equation}
    L_j
    =O\!\left(\log\!\left[\gamma_j n l(d)\right]\right),
    \qquad
    L_\Pi=\max_{0\le j\le d}L_j
    =O\!\left(\log\!\left[\gamma_{\max}n l(d)\right]\right),
\label{eq:app-explicit-loader-cost}
\end{equation}
where $\gamma_{\max}=\max_j\gamma_j$. This is a data-access and circuit-depth assumption rather than a generic state-preparation or total-gate bound. To prepare Eq.~\eqref{eq:app-purification}, first prepare $\gamma_k^{-1/2}\sum_\ell|\ell\rangle_R|\ell\rangle_P$ and then apply $V_{\Gamma,k}$ to the path register. Hence
\begin{equation}
    T_{\mathrm{prep}}^{\mathrm{dir}}
    =O(L_k)
    =O\!\left(\log\!\left[\gamma_k n l(d)\right]\right).
\label{eq:app-prep-direct}
\end{equation}

\subsection{Encoding and sparse access of the boundary operator}
\label{app:boundary-circuit}

The position-label encoding assigns the $l(d)$-qubit register defined in Eq.~\eqref{eq:app-path-register-size} to each vertex $j$.  For the path $p_k=v_0\cdots v_k$, the integer stored in the $j$-th register is
\begin{equation}
    a_j=
    \begin{cases}
       r+1, & j=v_r,\\
       0, & j\notin p_k.
    \end{cases}
\end{equation}
The complete path register contains $n\lceil\log(d+2)\rceil$ qubits.

To delete position $r\in\{0,\ldots,k\}$, every vertex register is updated reversibly according to
\begin{equation}
    a_j\longmapsto \bar a_j^{(r)}=
    \begin{cases}
       0, & a_j=r+1,\\
       a_j-1, & a_j>r+1,\\
       a_j, & 0\leq a_j<r+1.
    \end{cases}
\label{eq:app-label-deletion}
\end{equation}
The first line removes the vertex at position $r$, the second shifts every later path position one step to the left, and the third leaves earlier and absent vertices unchanged.  A phase controlled by the parity of $r$ supplies the coefficient $(-1)^r$.  Thus
\begin{equation}
    \partial_k|p_k\rangle
    =\sum_{r=0}^{k}(-1)^r|p_k^{\widehat r}\rangle.
\label{eq:app-boundary-action}
\end{equation}

The adjoint $\partial_k^\dagger$ reverses Eq.~\eqref{eq:app-label-deletion}.  Given a $(k-1)$-path, an absent vertex $u$, and an insertion position $r$, the circuit increases every label larger than $r$ by one and writes $r+1$ into the register of $u$.  The same parity phase gives $(-1)^r$.  All temporary comparison flags are uncomputed at the end.

A $k$-path has $k+1$ deletion neighbours.  A $(k-1)$-path has at most $(k+1)(n-k)$ insertion candidates.  Hence the row sparsity of
\begin{equation}
    B=\partial+\partial^\dagger
\end{equation}
is bounded by
\begin{equation}
    s_B\leq(d+1)+(d+1)n=O(nd).
\label{eq:app-sparsity}
\end{equation}
A sparse location oracle $O_{\mathrm{loc}}$ takes a path and a neighbour index and returns the corresponding deletion or insertion candidate.  A value oracle $O_{\mathrm{val}}$ returns $0$, $+1$, or $-1$, with the sign determined by the insertion and deletion position.  Counting one coherent directed-edge query as unit cost $O(1)$, scanning the $n$ registers, comparing and reversibly updating their $l(d)$-bit labels, and checking at most $O(d)$ directed edges gives
\begin{equation}
    T_B=O\!\left(nl(d)+d\right)
\label{eq:app-B-query-cost}
\end{equation}
for one use of the resulting sparse block-encoding.  Additional logarithmic factors from reversible arithmetic and finite-precision rotations are absorbed into the $\widetilde O$ notation used for the final runtime.

For completeness, the deletion part can also be block-encoded directly.  Introduce a position register $R$ and prepare
\begin{equation}
    W_k|0\rangle_R
    =\frac{1}{\sqrt{k+1}}\sum_{r=0}^{k}|r\rangle_R.
\end{equation}
For each $r$, let $U_r$ implement the signed reversible deletion
\begin{equation}
    U_r:\bigotimes_{j=1}^{n}|a_j\rangle_j|0\rangle_D
    \longmapsto
    (-1)^r\bigotimes_{j=1}^{n}|\bar a_j^{(r)}\rangle_j|j_r\rangle_D,
\end{equation}
where $j_r$ is the deleted vertex.  The register $D$ retains $j_r$ so that the update is reversible.  With
\begin{equation}
    |\chi_n\rangle_D=\frac{1}{\sqrt n}\sum_{j=1}^{n}|j\rangle_D,
\end{equation}
and with $V_{\partial_k}$ denoting preparation of $R$, the controlled $U_r$, and unpreparation of $R$, the selected block is
\begin{equation}
    (\langle0|_R\langle\chi_n|_D\otimes I)
    V_{\partial_k}
    (|0\rangle_R|0\rangle_D\otimes I)
    =\frac{\partial_k}{(k+1)\sqrt n}.
\label{eq:app-boundary-block}
\end{equation}
The reverse circuit block-encodes $\partial_k^\dagger$.  A one-qubit linear combination of these two circuits gives the Dirac block, and the direct sum over $k$ gives a unitary $U_B$ satisfying
\begin{equation}
    (\langle0^{a_B}|\otimes I)U_B(|0^{a_B}\rangle\otimes I)
    =\frac{B}{\alpha_B}.
\label{eq:app-B-block}
\end{equation}
For the standard sparse-matrix block-encoding with entries of magnitude at most one, $\alpha_B=O(s_B)=O(nd)$.

\subsection{Coherent projection and simulation of the embedded Hodge Laplacian}
\label{app:projected-simulation}

Define the easily recognized reference-label subspace
\begin{equation}
    Q_k=\sum_{\ell=0}^{\gamma_k-1}|\ell\rangle\langle\ell|.
\end{equation}
Equation~\eqref{eq:app-loader} maps this subspace onto $\Gamma_k$.  Therefore
\begin{equation}
    \Pi_{\Gamma,k}
    =V_{\Gamma,k}Q_kV_{\Gamma,k}^\dagger.
\label{eq:app-projector-loader}
\end{equation}
Operationally, the circuit unloads a path state to the simple label basis, checks whether the label lies in $0\leq\ell<\gamma_k$, and loads it back.

Let $R_{Q,k}=2Q_k-I$ be the reflection about valid labels.  Conjugating it with the loader gives
\begin{equation}
    R_{\Gamma,k}
    =V_{\Gamma,k}R_{Q,k}V_{\Gamma,k}^\dagger
    =2\Pi_{\Gamma,k}-I.
\label{eq:app-Gamma-reflection}
\end{equation}
An exact block-encoding of $\Pi_{\Gamma,k}$ is obtained with one additional qubit:
\begin{equation}
    U_{\Pi,k}
    =(H\otimes I)
    \left(
      |0\rangle\langle0|\otimes I
      +|1\rangle\langle1|\otimes R_{\Gamma,k}
    \right)
    (H\otimes I),
\end{equation}
for which
\begin{equation}
    (\langle0|\otimes I)U_{\Pi,k}(|0\rangle\otimes I)
    =\frac{I+R_{\Gamma,k}}{2}
    =\Pi_{\Gamma,k}.
\label{eq:app-projector-block}
\end{equation}
Taking the degree-controlled direct sum over $j$ yields a block-encoding $U_\Pi$ of $\Pi=\bigoplus_j\Pi_{\Gamma,j}$. One use of $U_\Pi$ consists of a controlled sector loader, a reversible range reflection on $0\leq\ell<\gamma_j$, and the controlled unloader. With Eq.~\eqref{eq:app-explicit-loader-cost}, its depth is
\begin{equation}
    T_\Pi^{\mathrm{dir}}
    =O(L_\Pi)
    =O\!\left(\log\!\left[\gamma_{\max} n l(d)\right]\right),
\label{eq:app-projector-cost-direct}
\end{equation}
where the range comparison contributes only a polylogarithmic term.
Under the standard-access promise in Eq.~\eqref{eq:app-standard-reflection-cost}, the degree-controlled direct sum of the sector reflections gives the total projector block-encoding with cost
\begin{equation}
    T_\Pi^{\mathrm{std}}
    =O\!\left(nl(d)+d\right).
\label{eq:app-projector-cost-standard}
\end{equation}

We now combine $U_\Pi$ and $U_B$.  A block-encoding product must use independent ancillary registers.  If $A_1,A_2,A_3$ denote the ancillas for the first projector, $U_B$, and the second projector, respectively, then the product circuit has selected block
\begin{align}
 &\bigl(\langle0|_{A_3}\langle0^{a_B}|_{A_2}\langle0|_{A_1}\otimes I\bigr)
 U_\Pi^{(A_3)}U_B^{(A_2)}U_\Pi^{(A_1)}
 \bigl(|0\rangle_{A_3}|0^{a_B}\rangle_{A_2}|0\rangle_{A_1}\otimes I\bigr)
 \nonumber\\
 &\hspace{4cm}=\frac{\Pi B\Pi}{\alpha_B}
 =\frac{B_\Gamma}{\alpha_B}.
\label{eq:app-BGamma-block}
\end{align}
Using independent ancillas prevents leakage from one selected block from returning to the selected subspace in the next multiplication.

Applying this product block-encoding twice, again with independent ancillas, gives
\begin{equation}
    (\langle0^{a_\Delta}|\otimes I)U_\Delta
    (|0^{a_\Delta}\rangle\otimes I)
    =\frac{B_\Gamma^2}{\alpha_B^2}
    =\frac{\Delta^\Gamma}{\alpha_B^2}.
\label{eq:app-Delta-block}
\end{equation}
Equivalently, one may multiply the five selected blocks in the order
\begin{equation}
    \Pi B\Pi B\Pi.
\end{equation}
The middle projector is essential: in general $\Pi B\Pi B\Pi\neq\Pi B^2\Pi$.

The identity with the Hodge Laplacian follows from the chain-complex structure.  On $\Gamma$,
\begin{equation}
    B_\Gamma=\partial^\Gamma+(\partial^\Gamma)^{*}.
\end{equation}
Since $(\partial^\Gamma)^2=0$ and $[(\partial^\Gamma)^{*}]^2=0$,
\begin{equation}
    B_\Gamma^2
    =(\partial^\Gamma)^{*}\partial^\Gamma
      +\partial^\Gamma(\partial^\Gamma)^{*}
    =\Delta^\Gamma.
\label{eq:app-dirac-square}
\end{equation}
Thus Eq.~\eqref{eq:app-Delta-block} is a block-encoding of the embedded Hodge Laplacian itself.

\subsubsection{Qubitization of the Laplacian block-encoding}

Set
\begin{equation}
    H_\Delta=\frac{\Delta^\Gamma}{\alpha_\Delta},
    \qquad
    \alpha_\Delta=\alpha_B^2,
    \qquad
    \|H_\Delta\|\leq1,
\label{eq:app-normalized-laplacian}
\end{equation}
so that Eq.~\eqref{eq:app-Delta-block} becomes
\begin{equation}
    (\langle0^{a_\Delta}|\otimes I)U_\Delta
    (|0^{a_\Delta}\rangle\otimes I)=H_\Delta.
\label{eq:app-Delta-selected-block}
\end{equation}
The $a_\Delta$ ancillary qubits identify the selected matrix block.  Define
\begin{equation}
    P_0=|0^{a_\Delta}\rangle\langle0^{a_\Delta}|\otimes I.
\label{eq:app-signal-projector}
\end{equation}
A general block-encoding $U_\Delta$ need not be Hermitian.  We therefore add one qubit $q$ and define the Hermitian unitary
\begin{equation}
    \mathcal U_\Delta
    =|0\rangle\langle1|_q\otimes U_\Delta
     +|1\rangle\langle0|_q\otimes U_\Delta^\dagger.
\label{eq:app-hermitianization}
\end{equation}
It satisfies
\begin{equation}
    \mathcal U_\Delta^\dagger=\mathcal U_\Delta,
    \qquad
    \mathcal U_\Delta^2=I.
\end{equation}
The enlarged signal projector and its reflection are
\begin{equation}
    \widetilde P=|+\rangle\langle+|_q\otimes P_0,
    \qquad
    \widetilde R=2\widetilde P-I.
\label{eq:app-qubitization-signal}
\end{equation}
Because $H_\Delta$ is Hermitian, the selected block of $\mathcal U_\Delta$ is still $H_\Delta$:
\begin{equation}
    \widetilde P\mathcal U_\Delta\widetilde P
    =H_\Delta
\end{equation}
inside the signal subspace.

Consider an eigenpair
\begin{equation}
    H_\Delta|\psi_j\rangle=x_j|\psi_j\rangle,
    \qquad
    |x_j|\leq1,
\label{eq:app-HDelta-eigenpair}
\end{equation}
and define the corresponding signal state
\begin{equation}
    |G_j\rangle
    =|+\rangle_q|0^{a_\Delta}\rangle|\psi_j\rangle.
\end{equation}
For $|x_j|<1$, define the normalized orthogonal state
\begin{equation}
    |B_j\rangle
    =\frac{\mathcal U_\Delta|G_j\rangle-x_j|G_j\rangle}
           {\sqrt{1-x_j^2}}.
\label{eq:app-bad-state}
\end{equation}
The two states span an invariant two-dimensional subspace.  In the ordered basis $\{|G_j\rangle,|B_j\rangle\}$,
\begin{equation}
    \mathcal U_\Delta
    \widehat{=}
    \begin{pmatrix}
        x_j & \sqrt{1-x_j^2}\\
        \sqrt{1-x_j^2} & -x_j
    \end{pmatrix},
    \qquad
    \widetilde R
    \widehat{=}
    \begin{pmatrix}
        1&0\\0&-1
    \end{pmatrix}.
\label{eq:app-two-dimensional-blocks}
\end{equation}
Here $\widehat{=}$ means the matrix representation restricted to that invariant subspace.  The limiting cases $x_j=\pm1$ follow by continuity.

The qubitized walk is
\begin{equation}
    W_\Delta=\widetilde R\mathcal U_\Delta.
\label{eq:app-qubitized-walk}
\end{equation}
Its restriction to the same subspace is
\begin{equation}
    W_\Delta
    \widehat{=}
    \begin{pmatrix}
        x_j & \sqrt{1-x_j^2}\\
        -\sqrt{1-x_j^2} & x_j
    \end{pmatrix}
    =
    \begin{pmatrix}
        \cos\theta_j & \sin\theta_j\\
        -\sin\theta_j & \cos\theta_j
    \end{pmatrix},
\label{eq:app-walk-rotation}
\end{equation}
where
\begin{equation}
    \cos\theta_j=x_j.
\end{equation}
Thus qubitization converts each unknown eigenvalue of $H_\Delta$ into the rotation angle, or equivalently the eigenphases $e^{\pm i\theta_j}$, of a quantum walk.  No eigenvalue is measured or supplied to the circuit in this step.

\subsubsection{Quantum signal processing and QSVT Hamiltonian simulation}

After qubitization, quantum signal processing transforms the walk rotation into the desired Hamiltonian-evolution phase.  A phase rotation about the signal reflection is
\begin{equation}
    S(\phi)=e^{i\phi\widetilde R}.
\label{eq:app-signal-rotation}
\end{equation}
Within the two-dimensional subspace above, it gives opposite phases to $|G_j\rangle$ and $|B_j\rangle$.  It can be implemented by a phase gate controlled on the signal ancillas being in the selected state.

Let
\begin{equation}
    \tau=\alpha_\Delta t=\alpha_B^2t.
\end{equation}
The classical preprocessing first constructs a polynomial $P_m(x)$ satisfying
\begin{equation}
    P_m(x)\approx e^{-i\tau x}
\end{equation}
uniformly on $[-1,1]$ with error at most $\epsilon_{\mathrm{sim}}$.  Equivalently, its even and odd parts approximate $\cos(\tau x)$ and $\sin(\tau x)$.  A classical QSP phase-synthesis algorithm then converts $P_m$ into fixed angles
\begin{equation}
    \boldsymbol\phi=(\phi_0,\phi_1,\ldots,\phi_m).
\end{equation}
This preprocessing depends only on $\tau$, the spectral interval $[-1,1]$, and the target error.  It does not require the eigenvalues $x_j$ or the eigenvectors $|\psi_j\rangle$.

The quantum circuit applies the walk and the signal rotations in the order
\begin{equation}
    \mathcal W_{\boldsymbol\phi}
    =S(\phi_0)W_\Delta S(\phi_1)W_\Delta
     \cdots W_\Delta S(\phi_m).
\label{eq:app-qsp-walk-sequence}
\end{equation}
For each invariant subspace, the synthesized phases make the selected matrix element approximate $P_m(x_j)$.  Therefore
\begin{equation}
    P_m(x_j)\approx e^{-i\tau x_j}.
\end{equation}
If $\nu_j$ is the corresponding eigenvalue of $\Delta^\Gamma$, then $x_j=\nu_j/\alpha_\Delta$ and
\begin{equation}
    e^{-i\tau x_j}
    =e^{-i\alpha_\Delta t(\nu_j/\alpha_\Delta)}
    =e^{-i\nu_jt}.
\end{equation}
By linearity, the same circuit gives the correct phase to every eigencomponent of an arbitrary input state and hence implements
\begin{equation}
    \mathcal W_{\boldsymbol\phi}
    \approx e^{-i\Delta^\Gamma t}
\label{eq:app-qsvt-evolution}
\end{equation}
on the signal subspace.  This procedure is qubitization followed by quantum signal processing.  Equivalently, it is the Hermitian-matrix instance of the quantum singular value transformation framework.  The block-encoding ancillas are not measured or postselected after each query; they are used coherently throughout the complete sequence~\cite{PhysRevLett.118.010501,Low2019hamiltonian,10.1145/3313276.3316366}.

The number of calls to $U_\Delta$ and $U_\Delta^\dagger$ is
\begin{equation}
    O\!\left(\alpha_\Delta|t|+\log\frac{1}{\epsilon_{\mathrm{sim}}}\right)
    =O\!\left(\alpha_B^2|t|+\log\frac{1}{\epsilon_{\mathrm{sim}}}\right).
\label{eq:app-qsvt-query}
\end{equation}
An alternative implementation avoids explicitly constructing $U_\Delta$.  Starting from the block-encoding of $B_\Gamma/\alpha_B$, QSVT can directly approximate
\begin{equation}
    f_B(x)=e^{-i\alpha_B^2tx^2}.
\end{equation}
For an eigenvalue $\mu$ of $B_\Gamma$,
\begin{equation}
    f_B(\mu/\alpha_B)=e^{-i\mu^2t},
\end{equation}
which is the phase generated by $\Delta^\Gamma=B_\Gamma^2$.

For the Betti-number algorithm, it is cheaper to perform phase estimation directly on the total projected Dirac operator $B_\Gamma$. Define
\begin{equation}
    g_k=\sqrt{\lambda_{\min}^{+}(\Delta_k^\Gamma)}.
\end{equation}
When the positive spectrum is empty, the entire $k$-sector is harmonic and no spectral-resolution step is needed. Otherwise, because the input is supported on $\Gamma_k$ and $B_\Gamma^2=\Delta^\Gamma$, every nonzero Dirac spectral component visible from that input has magnitude at least $g_k$. Resolving zero from the nonzero spectrum therefore requires simulated time $O(g_k^{-1})$. Since the block-encoding normalization is $\alpha_B$, the number of block-encoding queries is
\begin{equation}
    \widetilde O(\alpha_B/g_k)
    =\widetilde O(\kappa_k),
    \qquad
    \kappa_k=\frac{\alpha_B}{g_k}.
\label{eq:app-kappa}
\end{equation}
The same-degree compression $\Pi_{\Gamma,k}B\Pi_{\Gamma,k}$ is not used: it is identically zero because $B$ changes degree by one. A finite-resolution implementation classifies $|\widetilde\mu|<g_k/2$ as zero, with phase-estimation error chosen below $g_k/4$.

\subsection{Derivation of the runtime bounds}

One use of the block-encoding of
\begin{equation}
    B_\Gamma=\Pi B\Pi
\end{equation}
contains one ambient Dirac block-encoding and two total-projector block-encodings. The constant factor two is irrelevant asymptotically. From Eqs.~\eqref{eq:app-sparsity} and \eqref{eq:app-B-query-cost},
\begin{equation}
    \alpha_B=O(nd),
    \qquad
    T_B=O\!\left(nl(d)+d\right).
\label{eq:app-explicit-B-resources}
\end{equation}
With
\begin{equation}
    g_k=\sqrt{\lambda_{\min}^{+}(\Delta_k^\Gamma)},
\end{equation}
for a nonempty positive spectrum; otherwise the spectral-resolution cost is absent. Then the smallest nonzero Dirac magnitude visible from an input in $\Gamma_k$, after block-encoding normalization, is $g_k/\alpha_B$. Resolving zero from the nonzero spectrum therefore requires
\begin{equation}
    N_{\mathrm{qry}}
    =\widetilde O\!\left(\frac{\alpha_B}{g_k}\right)
    =\widetilde O\!\left(\frac{nd}{g_k}\right)
\label{eq:app-explicit-query-number}
\end{equation}
queries to the total projected-Dirac block-encoding. This is the parameter $\widetilde O(\kappa_k)$ with $\kappa_k=\alpha_B/g_k$. Explicitly simulating the squared operator would instead introduce the squared normalization and spectral scale.

Each run produces the outcome zero with probability
\begin{equation}
    c_k=\frac{\beta_k}{\gamma_k}.
\end{equation}
Estimating this probability to additive error $\varepsilon$ with failure probability at most $\delta$ requires
\begin{equation}
    M=O\!\left(\frac{\log(1/\delta)}{\varepsilon^2}\right)
\label{eq:app-sampling}
\end{equation}
independent repetitions.  The total runtime is therefore the number of repetitions times the sum of the state-preparation cost and the cost of one phase-estimation run.

Under the standard-access promise in Eq.~\eqref{eq:app-standard-reflection-cost}, one projected-Dirac query costs
\begin{equation}
    O\!\left(nl(d)+d\right),
\end{equation}
because the ambient Dirac circuit and each of the two total-projector block-encodings have this same asymptotic cost.  Combining Eqs.~\eqref{eq:app-prep-standard}, \eqref{eq:app-explicit-query-number}, and \eqref{eq:app-sampling} gives
\begin{equation}
    T_{\mathrm{tot}}^{\mathrm{std}}
    =\widetilde O\!\left[
      \left(nl(d)+d\right)
      \left(
        \zeta_k^{-1/2}+\frac{nd}{g_k}
      \right)
      \frac{\log(1/\delta)}{\varepsilon^2}
    \right].
\label{eq:app-total-standard}
\end{equation}
The first term is amplitude-amplified state preparation; the second is the cost of the projected-Dirac queries needed to resolve the spectral gap.

Under direct path-space access, Eq.~\eqref{eq:app-explicit-loader-cost} gives
\begin{equation}
    L_k=O\!\left(\log\!\left[\gamma_k n l(d)\right]\right),
    \qquad
    L_\Pi=O\!\left(\log\!\left[\gamma_{\max} n l(d)\right]\right).
\end{equation}
The input state costs $O(L_k)$ to prepare, while one query to the total projected Dirac costs
\begin{equation}
    O\!\left(nl(d)+d+L_\Pi\right).
\end{equation}
Consequently,
\begin{equation}
\begin{aligned}
    T_{\mathrm{tot}}^{\mathrm{dir}}
    =\widetilde O\!\Bigg[
      \Bigg(&
        \log\!\left[\gamma_k n l(d)\right]
        +\left(
          nl(d)+d
          +\log\!\left[\gamma_{\max} n l(d)\right]
        \right)\frac{nd}{g_k}
      \Bigg)
      \frac{\log(1/\delta)}{\varepsilon^2}
    \Bigg].
\end{aligned}
\label{eq:app-total-direct}
\end{equation}
Here the first logarithmic term is direct preparation of the $k$-sector purification, and the second line is the ambient Dirac cost plus degree-controlled projector access, repeated $\widetilde O(nd/g_k)$ times.

The notation $\widetilde O$ hides polylogarithmic factors from reversible arithmetic, finite-precision Hamiltonian simulation, and phase estimation, but it does not hide the explicitly displayed sampling factor $\varepsilon^{-2}$.  Equations~\eqref{eq:app-total-standard} and \eqref{eq:app-total-direct} are the detailed derivations of Eqs.~\eqref{eq:main-complexity-standard} and \eqref{eq:main-complexity-direct}.

These expressions combine oracle-call counts with circuit depth per call. In particular, $L_j$ and $L_\Pi$ are depths under a parallel coherent-loading architecture. They are not total gate counts and do not include the classical memory, routing network, or parallel hardware needed to store the path data. Those resources must be reported separately for a concrete implementation.

\subsection{Normalized and exact Betti numbers}

The efficient statistical output is an additive estimate of
\begin{equation}
    c_k=\frac{\beta_k}{\gamma_k}.
\end{equation}
The natural estimator is $\widehat\beta_k=\gamma_k\widehat c_k$. To recover the exact integer $\beta_k$, one must know $\gamma_k$ and obtain an estimate satisfying
\begin{equation}
    |\widehat c_k-c_k|<\frac{1}{2\gamma_k},
    \qquad
    \beta_k=\operatorname{round}(\gamma_k\widehat c_k).
\end{equation}
If $\gamma_k$ is exponential, this can require exponentially small additive error and correspondingly many repetitions.
The complexity claims should therefore be interpreted as normalized Betti-number estimation unless an additional promise makes exact recovery efficient.

\subsection{Consequences for the speedup claim}

The runtime separates into four resources: state preparation, operator access, spectral resolution, and statistical precision.  Under standard digraph access, $\zeta_k^{-1/2}$ can be exponentially large, and the explicit bound also assumes the coherent $\Gamma_k$-reflection promise in Eq.~\eqref{eq:app-standard-reflection-cost}; hence no unconditional exponential speedup is claimed.  Under direct path-space access, the Grover factor is replaced by the logarithmic loader depth $O(\log[\gamma_k n l(d)])$ under the parallel-loading assumption.  If $g_k^{-1}$ is polynomially bounded, the resulting quantum runtime is polynomial in the assumed path-space description and may be exponentially smaller than classical path-homology algorithms in regimes where the path length grows.  For fixed small $k$, the best-known classical complexity is polynomial in $n$, and the corresponding advantage is at most polynomial.

\section{Examples of Classical Computation of GLMY Homology}
\label{examples_classical_computation_GLMY}

In this section, we present a direct computation of the chain groups, boundary maps in the sequence of Eq.~\eqref{simplicial_chain_G} and the corresponding Betti numbers.
\begin{align}\label{simplicial_chain_G}
    \cdots\stackrel{\partial^{\Gamma}_5}{\longrightarrow}\Gamma_4 \stackrel{\partial^{\Gamma}_4}{\longrightarrow} \Gamma_{3} \stackrel{\partial^{\Gamma}_{3}}{\longrightarrow} \Gamma_{2}\stackrel{\partial^{\Gamma}_{2}}{\longrightarrow} \Gamma_1 \stackrel{\partial^{\Gamma}_{1}}{\longrightarrow} \Gamma_0 \stackrel{\partial^{\Gamma}_{0}}{\longrightarrow} 0.
\end{align}

\begin{example}
Let $G$ be the digraph shown in Fig.~$2$ (a).

We will show that all its Betti numbers are zero except for $\beta_0$. It is clear that the groups of allowed paths are as follows:
\begin{equation}
    \begin{aligned}
        \Acal_0&=&\spn\{&1,2,3,4\},\\
        \Acal_1&=&\spn\{&12,13,14,23,24,34\}\\
        \Acal_2&=&\spn\{&123,124,134,234\}\\
        \Acal_3&=&\spn\{&1234\}\\
    \end{aligned}
\end{equation}
For $k\geq 4$, we have $\Acal_k=0$. 
The $k$-th chain groups $\Gamma_k=\Acal_k=0$ and $k$-th boundary maps $\partial_k=\partial_k^{\Gamma}=0$. 
When $k=0$, the chain group $\Gamma_0$ is always equal to the group of all vertices, i.e., $\Gamma_0=\Acal_0=\spn\{1,2,3,4\}$. The norm matrix of $\langle-,-\rangle_{\Gamma_0}$ is $\Id_{4\times 4}$. And the $0$-th boundary map $\partial_0=\partial^{\Gamma}_0=0$.

For $1\leq k\leq 3$, from the definition, the boundary operators are listed as follows
\begin{equation}
    \begin{aligned}
\partial_3(1234)&=&(234-134+124-123)\\
\partial_2(123,124,134,234)&=&(23-13+12,24-14+12,34-14+13,34-24+23)\\
 \partial_1(12,13,14,23,24,34)&=&(2-1,3-1,4-1,3-2,4-2,4-3)
    \end{aligned}
\end{equation}
Therefore, the third chain group $\Gamma_3=\Acal_3+\partial_4\Acal_4=\spn\{1234\}$. The norm matrix of $\langle-,-\rangle_{\Gamma_3}$ is $N_3=1$.
The second chain group $\Gamma_2=\Acal_2+\partial_3\Acal_3=\spn\{123,124,134,234\}=\Acal_2$. The norm matrix of $\langle-,-\rangle_{\Gamma_2}$ is $N_2=\Id_{4\times 4}$. And the first chain group $\Gamma_1=\Acal_1+\partial_2\Acal_2=\spn\{12,13,14,23,24,34\}=\Acal_1$ and the norm matrix of $\langle-,-\rangle_{\Gamma_1}$ is $N_1=\Id_{6\times 6}$.

Then the boundary matrices $D_k^{\Gamma}$ of the boundary maps $\partial^{\Gamma}_k$ when $k=1,2,3$ are listed as following
\begin{equation}
    \begin{aligned}
        \partial^{\Gamma}_3&(1234)\\
        =&(123,124,134,234)D_3^{\Gamma}\\
        =&(123,124,134,234)
        \begin{pmatrix}
           -1\\
           1\\
           -1\\
           1\\
   \end{pmatrix}\\
\partial^{\Gamma}_2&(123,124,134,234)\\
        =&(12,13,14,23,24,34)D_2^{\Gamma}\\
        =&(12,13,14,23,24,34)\begin{pmatrix}
             1 & 1 & 0 & 0 \\
             -1& 0 & 1 & 0\\
             0 & -1& -1& 0 \\
             1 & 0 & 0 & 1 \\
             0 & 1 & 0 &-1 \\
             0 & 0 & 1 & 1 \\
        \end{pmatrix}\\
        \partial^{\Gamma}_1&(12,13,14,23,24,34)\\
        =&(1,2,3,4)D_1^{\Gamma}\\
        =&(1,2,3,4)\begin{pmatrix}
            -1& -1& -1& 0 & 0 & 0\\
            1 & 0 & 0 & -1& -1& 0\\
            0 & 1 & 0 & 1 & 0 & -1\\
            0 & 0 & 1 & 0 & 1 & 1\\ 
        \end{pmatrix}
    \end{aligned}
\end{equation}

Since all the norm matrix are identity, computation shows that the Hodge-Laplacian operators are as the equation~\eqref{eq:delta-G-eg1}. It is clear that $\Delta^{\Gamma}_3,\ \Delta^{\Gamma}_2$ and $\Delta^{\Gamma}_1$ are full rank while $\rk \Delta^{\Gamma}_0=3$. Hence the Betti numbers are $\beta_0=1$ and the others are 0.
\begin{equation}\label{eq:delta-G-eg1}
    \begin{aligned}
        &\Delta^{\Gamma}_3&=&(D_3^{\Gamma})^{\dagger}D_3^{\Gamma}&=& 4\Id_{1\times 1}\\
        &\Delta^{\Gamma}_2&=&D_3^{\Gamma}(D_3^{\Gamma})^{\dagger}+(D_2^{\Gamma})^{\dagger}D_2^{\Gamma}&=&4\Id_{4\times 4}\\
        &\Delta^{\Gamma}_1&=&D_2^{\Gamma}(D_2^{\Gamma})^{\dagger}+(D_1^{\Gamma})^{\dagger}D_1^{\Gamma}&=&4\Id_{6\times 6}\\
        &\Delta^{\Gamma}_0&=&D_1^{\Gamma}(D_1^{\Gamma})^{\dagger}&=&\begin{pmatrix}
            3 &-1&-1&-1\\  
            -1 &3&-1&-1\\
            -1&-1&3&-1\\
            -1&-1&-1&3\\ 
        \end{pmatrix}\\
    \end{aligned}
\end{equation}
\end{example}

\begin{figure*}\label{fig:digraph-eg}
    \subfigure[]{
        \includegraphics[width=5cm]{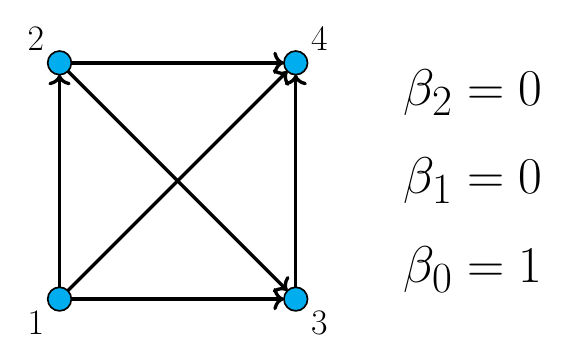}}
\hspace{20mm}
    \subfigure[]{
\includegraphics[width=5cm]{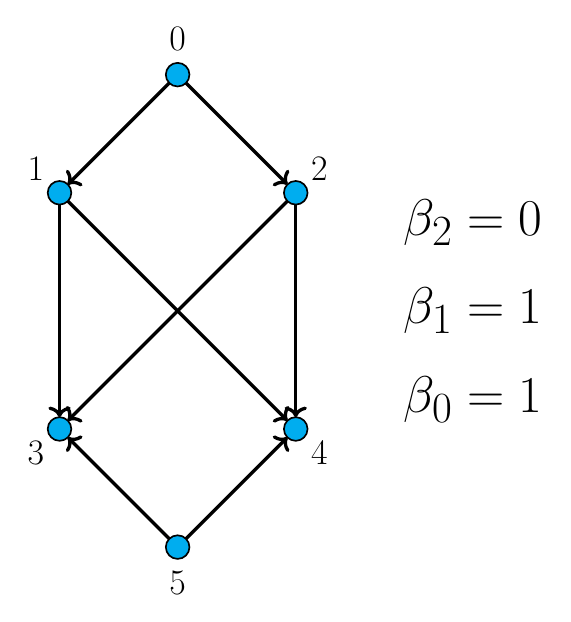}
        }
\caption{Two examples of digraphs.}
\end{figure*}

\begin{example}
Let $G$ be the digraph as the Fig.~$2$ (b).

It was proved in \cite{grigor2012homologies} that this digraph satisfies $\beta_0=\beta_1=1$ and $\beta_k=0$ for any $k\geq 2$. We verify these properties by direct computation.
\begin{equation}
    \begin{aligned}
        \Acal_0&=&\spn\{&0,1,2,3,4,5\},\\
        \Acal_1&=&\spn\{&01,02,13,14,23,24,53,54\}\\
        \Acal_2&=&\spn\{&013,014,023,024\}\\
    \end{aligned}
\end{equation}
And for $k\geq 3$, the group of allowed paths $\Acal_k=0$ all vanish.
The $k$-th chain groups $\Gamma_k=\Acal_k=0$ and $k$-th boundary maps $\partial_k=\partial_k^{\Gamma}=0$. 
When $k=0$, the chain group $\Gamma_0$ is always equal to the group of all vertices, i.e., $\Gamma_0=\Acal_0=\spn\{0,1,2,3,4,5\}$. The norm matrix of $\langle-,-\rangle_{\Gamma_0}$ is $\Id_{6\times 6}$. And the $0$-th boundary map $\partial_0=\partial^{\Gamma}_0=0$.

When $k=1,2$, the boundary maps are given as
\begin{equation}
    \begin{aligned}
        \partial_2(013,014,023,024)&=&(13-03+01,14-04+01,23-03+02,24-04+02)\\
        \partial_1(01,02,13,14,23,24,53,54,03,04)&=&(1-0,2-0,3-1,4-1,3-2,4-2,3-5,4-5,3-0,4-0)
    \end{aligned}    
\end{equation}
Hence the second chain group $\Gamma_2=\Acal_2+\partial_3\Acal_3=\spn\{013,014,023,024\}$. The norm matrix of $\langle-,-\rangle_{\Gamma_2}$ is $N_2=\Id_{4\times 4}$. The first chain group $\Gamma_1=\Acal_1+\partial_2\Acal_2=\spn\{01,02,13,14,23,24,53,54,03,04\}$, where $03,04$ are generated from $\partial_2\Acal_2$. The norm matrix of $\langle-,-\rangle_{\Gamma_1}$ is $N_1=\Id_{10\times 10}$.

The computation shows that the boundary matrices $D_2^{\Gamma}$ and $D_1^{\Gamma}$ of the boundary maps $\partial^{\Gamma}_2$ and $\partial^{\Gamma}_1$, respectively, are as follows.
\begin{equation}
    \begin{aligned}
        \partial^{\Gamma}_2&(013,014,023,024)\\
        =&(01,02,13,14,23,24,53,54,03,04)D_2^{\Gamma}\\
        =&(01,02,13,14,23,24,53,54,03,04)
        \begin{pmatrix}
            1 & 1 & 0 & 0\\
            0 & 0 & 1 & 1\\
            1 & 0 & 0 & 0\\
            0 & 1 & 0 & 0\\
            0 & 0 & 1 & 0\\
            0 & 0 & 0 & 1\\
            0 & 0 & 0 & 0\\
            0 & 0 & 0 & 0\\
            -1& 0 & -1& 0\\
            0 & -1& 0 & -1\\
   \end{pmatrix}\\
           \partial^{\Gamma}_1&(01,02,13,14,23,24,53,54,03,04)\\
        =&(0,1,2,3,4,5)D_1^{\Gamma}\\
        =&(0,1,2,3,4,5)\left(
        \begin{array}{cccccccccc}
            -1& -1& 0 & 0 & 0 & 0 & 0 & 0 & -1& -1  \\
            1 & 0 & -1& -1& 0 & 0 & 0 & 0 & 0 & 0 \\
            0 & 1 & 0 & 0 & -1& -1& 0 & 0 & 0 & 0 \\
            0 & 0 & 1 & 0 & 1 & 0 & 1 & 0 & 1 & 0\\
            0 & 0 & 0 & 1 & 0 & 1 & 0 & 1 & 0 & 1 \\
            0 & 0 & 0 & 0 & 0 & 0 & -1& -1& 0 & 0 \\
        \end{array}
            \right)
    \end{aligned}
\end{equation}

Finally the Hodge-Laplacian operators are given in Eq.~\eqref{eq:delta-G-eg2}. The Betti numbers are $\beta_0=\beta_1=1$, where $\ker\Delta^{\Gamma}_1$ is generated by $(-13+14-23+24+3\,53-3\,54-03+04)$ and $\ker\Delta^{\Gamma}_0$ is generated by $(0+1+2+3+4+5)$. All remaining Betti numbers vanish.
\begin{equation}\label{eq:delta-G-eg2}
    \begin{aligned}
        &\Delta^{\Gamma}_2&=&(D_2^{\Gamma})^{\dagger}D_2^{\Gamma}&=&\left(
        \begin{array}{cccc}
3& 1& 1& 0\\1& 3& 0& 1\\1& 0& 3& 1\\0& 1& 1& 3\\
        \end{array}
            \right)\\
        &\Delta^{\Gamma}_1&=&D_2^{\Gamma}(D_2^{\Gamma})^{\dagger}+(D_1^{\Gamma})^{\dagger}D_1^{\Gamma}&=&\left(
        \begin{array}{cccccccccc}
4& 1& 0& 0& 0& 0& 0& 0& 0& 0\\1& 4& 0& 0& 0& 0& 0& 0& 0& 0\\0& 0& 3& 1& 1& 0& 1& 0& 0& 0\\0& 0& 1& 3& 0& 1& 0& 1& 0& 0\\0& 0& 1& 0& 3& 1& 1& 0& 0& 0\\0& 0& 0& 1& 1& 3& 0& 1& 0& 0\\0& 0& 1& 0& 1& 0& 2& 1& 1& 0\\0& 0& 0& 1& 0& 1& 1& 2& 0& 1\\0& 0& 0& 0& 0& 0& 1& 0& 4& 1\\0& 0& 0& 0& 0& 0& 0& 1& 1& 4\\
        \end{array}
            \right)\\
        &\Delta^{\Gamma}_0&=&D_1^{\Gamma}(D_1^{\Gamma})^{\dagger}&=&\left(
        \begin{array}{cccccc}
4& -1& -1& -1& -1& 0\\-1& 3& 0& -1& -1& 0\\-1& 0& 3& -1& -1& 0\\-1& -1& -1& 4& 0& -1\\-1& -1& -1& 0& 4& -1\\0& 0& 0& -1& -1& 2\\
        \end{array}
            \right)\\
    \end{aligned}
\end{equation}
\end{example}

\bibliography{Reference.bib}
\widetext

\end{document}